\documentclass[fleqn,usenatbib]{mnras}

\usepackage{newtxtext,newtxmath}
\usepackage[T1]{fontenc}

\DeclareRobustCommand{\VAN}[3]{#2}
\let\VANthebibliography\thebibliography
\def\thebibliography{\DeclareRobustCommand{\VAN}[3]{##3}\VANthebibliography}

\newcommand{\vs}{$v\sin i$}
\newcommand{\kms}{km\,s$^{-1}$}
\newcommand{\ms}{m\,s$^{-1}$}

\usepackage{graphicx}	% Including figure files
\usepackage{amsmath}	% Advanced maths commands

\title[Doppler Imaging of the Sun]{Using Doppler Imaging to model stellar activity and search for planets around Sun-like stars}

\author[B. Klein et al.]{
Baptiste Klein,$^{1}$\thanks{E-mail: baptiste.klein@physics.ox.ac.uk}
Suzanne Aigrain,$^{1}$
Michael Cretignier,$^{1}$
Xavier Dumusque,$^{2}$
Khaled Al Moulla,$^{3}$
\newauthor
Jean-François Donati,$^{4}$
Niamh K. O'Sullivan,$^{1}$
Haochuan Yu,$^{1}$
Andrew Collier Cameron,$^{5}$
\newauthor
Oscar Barrag\'an,$^{1,6}$
Annelies Mortier,$^{7}$
Alessandro Sozzetti$^{8}$
\\
$^{1}$Department of Physics, University of Oxford, Oxford OX1 3RH, UK\\
$^{2}$Observatoire Astronomique de l’Université de Genève, Chemin Pegasi 51, Versoix 1290, Switzerland\\
$^{3}$Instituto de Astrofísica e Ciências do Espaço, Universidade do Porto, CAUP, Rua das Estrelas, 4150-762 Porto, Portugal\\
$^{4}$CNRS, IRAP, Université de Toulouse, 14 avenue Belin, F-31400 Toulouse, France\\
$^{5}$Centre for Exoplanet Science / SUPA, School of Physics \& Astronomy, University of St Andrews, North Haugh ST ANDREWS, Fife KY16 9SS, UK\\
$^{6}$Department of Physics, University of Warwick, Coventry CV4 7AL, UK\\
$^{7}$School of Physics \& Astronomy, University of Birmingham, Edgbaston, Birmingham, B15 2TT, UK \\
$^{8}$INAF – Osservatorio Astrofisico di Torino, via Osservatorio 20, Pino Torinese 10025, Italy
}

\date{Accepted 2025 August 07. Received 2025 August 06; in original form 2025 May 13}

\pubyear{2025}

\begin{document}
\label{firstpage}
\pagerange{\pageref{firstpage}--\pageref{lastpage}}
\maketitle

\begin{abstract}
Doppler Imaging (DI) is a well-established technique to map a physical field at a stellar surface from a time series of high-resolution spectra. In this proof-of-concept study, we aim to show that traditional DI algorithms, originally designed for rapidly-rotating stars, have also the ability to model the activity of Sun-like stars, when observed with new-generation highly-stable spectrographs, and search for low-mass planets around them. We used DI to retrieve the relative brightness distribution at the surface of the Sun from radial velocity (RV) observations collected by HARPS-N between 2022 and 2024. The brightness maps obtained with DI have a typical angular resolution of $\sim$36$^{\circ}$ and are a good match to low-resolution disc-resolved Dopplergrams of the Sun at epochs when the absolute, disc-integrated RV exceeds $\sim$2\,\ms. The RV residuals after DI correction exhibit a dispersion of about 0.6\,\ms, comparable with existing state-of-the-art activity correction techniques. Using planet injection-recovery tests, we also show that DI can be a powerful tool for blind planet searches, so long as the orbital period is larger than $\sim$100\ days (i.e. 3 to 4 stellar rotation periods), and that it yields planetary mass estimates with an accuracy comparable to, for example, multi-dimensional Gaussian process regression. Finally, we highlight some limitations of traditional DI algorithms, which should be addressed to make DI a reliable alternative to state-of-the-art RV-based planet search techniques.
\end{abstract}


%%%%%%%%%%%%%%%%%%%%%%%%%%%%%%%%%%%%%%%%%%%%%%%%%%
%%%%%%%%%%%%%%%%% BODY OF PAPER %%%%%%%%%%%%%%%%%%

\section{Introduction}

The detection and characterisation of temperate Earth-mass planets around Sun-like stars is one of the most exciting and challenging prospects in contemporary astrophysics. New-generation extreme-precision spectrographs like ESPRESSO \citep{pepe2021,faria2022}, EXPRES \citep{petersburg2020,Blackman2020,brewer2020}
and NEID \citep{schwab2016,gupta2025}, routinely deliver radial velocities (RVs) with sub-meter-per-second precision. With ongoing and forthcoming long-baseline planet search programs, such as the PLATO mission \citep[including its RV follow-up;][]{rauer2014}, the NEID Earth Twin Survey \citep{gupta2021,gupta2025}, or the Terra Hunting Experiment \citep{thompson2016,hall2018}, the detection of an Earth twin in the solar neighbourhood seems to be within reach.

Stellar variability is the main obstacle to such a detection \citep{fischer2016,crass2021,meunier2023}. Photospheric flows \citep[e.g. p-mode oscillations, supergranulation, meridional circulation; see][]{kjeldsen1995,dumusque2011,meunier2015,meunier2019,chaplin2019,osullivan2024,osullivan2025} and magnetic activity \citep[e.g. magnetically-active regions, magnetic cycles; see][]{saar1997,desort2007,meunier2010,dumusque2011c,meunier2017} distort the stellar absorption line profiles, giving rise to RV signals that mask planet signatures.

%%% Time-domain methods and limitations
The modelling of stellar activity RV signals has become a very active field of research in the last decade \citep[see][]{zhao2022}. These signals are often modelled in the time domain with Gaussian Processes \citep[GPs; see][for a review]{aigrain2022}. In particular, multi-dimensional GPs, where RV signals are modelled jointly with activity proxies, provide a trackable, robust and flexible stellar activity modelling framework, often considered as state of the art in the community \citep[e.g.][]{rajpaul2015,delisle2022,barragan2022,barragan2023,hara2025}. Despite their popularity, GPs suffer from key limitations in the search for low-mass long-period planets. In general, current GP models lack the flexibility to capture the RV signals of stars with rapidly-evolving activity (i.e. where active regions evolve on time scales similar to the star's rotation period), which can lead to inaccuracies in the estimates of the planet parameters \citep[e.g.][]{blunt2023}. Moreover, magnetic cycles induce significant variations in the properties of stellar activity signals \citep[e.g. rotation period and evolution timescale, due to changes in the filling factor of active regions;][]{klein2024b}, which are generally not accounted for in GP covariance functions.

%%% Wavelength-domain methods 
To overcome these limitations, the community is increasingly focusing on wavelength-domain activity filtering techniques. Approaches exploiting activity-induced changes in the shape of spectral lines have been shown to robustly filter cycle-induced activity variations, increasing the RV sensitivity to long-period planet signatures \citep[e.g.][]{jones2017,cameron2021,deBeurs2022,wilson2022,john2022,john2023,cretignier2023,klein2024b,zhao2024,yu2024}. On the other hand, by leveraging the wealth of information delivered by new-generation \'echelle spectrographs, line-by-line techniques enable a more detailed physical characterisation of stellar activity, which is known to affect different spectral lines in distinct ways \citep[e.g.][]{thompson2017,dumusque2018,cretignier2020,cretignier2022,bellotti2022,cretignier2022,almoulla2022a,lienhard2022,lienhard2023,almoulla2024,rescigno2025}.

%%% Doppler Imaging 
In this study, we investigate how Doppler Imaging \citep[see][]{vogt1987,donati2009,kochukhov2016} can be used to mitigate stellar activity and search for planet signatures in the spectra of Sun-like stars. This physically-motivated technique aims at inverting a time series of high-resolution spectra into a distribution of a physical field at the stellar surface (e.g. relative brightness, temperature, magnetic field, chemical composition). Historically, the mapping of brightness inhomogeneities from intensity spectra, which the present study focuses on, has been limited to rapidly-rotating stars such as giants and sub-giants \citep[e.g.][]{donati1999,strassmeier1999,rottenbacher2017}, T Tauri stars \citep[e.g.][]{cameron1994,hatzes1995,donati2000,donati2012,donati2014,yu2019,finociety2021}, stars in close-in binary systems \citep[e.g.][]{zaire2021,zaire2022} and even brown dwarfs \citep{crossfield2014,luger2021}. New-generation highly-stable high-resolution spectrographs have the ability to resolve the profile distortions induced by active regions at the surface of more slowly rotating stars \citep[as already showcased in][]{donati1995,hebrard2016}, which suggests that Doppler Imaging could be applied to Sun-like stars, provided that the signal-to-noise ratio (SNR) of the spectra is high enough, and that the temporal sampling is sufficiently dense.

%%% Plan of the paper
This proof-of-concept study aims to show that traditional Doppler Imaging can not only be applied to Sun-like stars, but also provide a reliable alternative to time-domain activity modelling techniques in the search for low-mass planetary signatures. In Section~\ref{sec:method}, we give a short review of Doppler Imaging techniques in stellar physics and describe our framework to apply it to Sun-like stars. In Section~\ref{sec:data}, we describe the HARPS-N Sun-as-a-star observations used in this case study, before modelling them with Doppler Imaging and comparing the results to resolved images of the Sun in Section~\ref{sec:Results}. Finally, we assess the sensitivity of Doppler Imaging to planet signatures in Section~\ref{sec:planets}, before discussing the current limitations of our framework and avenues to address them in Section~\ref{sec:conclusion}.

\section{Method} \label{sec:method}

\subsection{Doppler Imaging in Stellar Physics}

Doppler Imaging (DI) is a long-standing method to map inhomogeneous structures on stellar surfaces from high-resolution spectra. DI has been historically used for the chemical mapping of Ap and Bp stars, which exhibit strong chemically peculiar absorption lines with periodic variations \citep[e.g.][]{Goncharskij1983,vogt1983,Khokhlov1986,vogt1987,hatzes1991,kochukhov2004,kochukhov2007}. Zeeman-Doppler Imaging (ZDI), an extension of DI to map stellar magnetic field topologies from polarized spectra, is one of the most succesful stellar imaging techniques \citep[see][for a review]{donati2009,kochukhov2016}. ZDI has been applied to a very wide range of stars, including massive early-type stars \citep[e.g.][]{donati2002,donati2006b}, chemically-peculiar stars \citep[e.g.][]{landstreet2000,bagnulo2002,kochukhov2010b}, RS CVn stars \citep[e.g.][]{donati1999,donati2003,kochukhov2013}, classical and weak-line T Tauri stars \citep[e.g.][]{donati2008,donati2012,folsom2016,yu2017,yu2019}, Sun-like stars \citep{petit2008,marsden2014,folsom2020,petit2021} and M dwarfs \citep[e.g.][]{morin2008,morin2010,klein2021a,lehmann2024}. The output magnetic topologies are excellent showcases of the dynamo processes in stellar interiors \citep[e.g.][]{donati2003,gastine2013,brun2017}, and the evolution of the field's properties with time is one of the most direct way to unambigously unveil magnetic cycles \citep[e.g.][]{fares2009,boro-saika2016,boro-saika2018,lavail2018,lehmann2021,bellotti2025}. Moreover, magnetic topologies are essential ingredients for modelling magnetised winds and their interaction with close-in planets \citep[e.g.][]{kavanagh2021,callingham2024,strugarek2025}.

In this study, we focus on the mapping of brightness inhomogeneities from intensity spectra. This method has historically been effective with rapidly-rotating active low-mass stars, such as classical and weak-line T Tauri stars, members of close-in binary systems (e.g. RS-CVn stars) and rapidly rotating giant stars (e.g. K Com stars). In contrast with temperature mapping, which uses a small number of spectral lines to constrain the spot properties \citep[see][]{berdyugina2005,strassmeier2009,afram2015}, brightness reconstructions focus on retrieving a fractional spot coverage, losing the absolute information on the spot temperature, but allowing one to combine thousands of spectral lines at the same time \citep[e.g. through Least Squares Deconvolution;][]{donati1997c}. Mapping the star's relative brightness is a way of modelling stellar activity, bypassing the RV computation process. In fact, DI has been used to jointly model the activity of pre-main-sequence stars whilst searching for planetary signatures in the recent litterature \citep[e.g.][]{petit2015,donati2017,yu2017,klein2021,klein2022}.

Doppler Imaging is an ill-posed inverse problem, which means that multiple solutions can fit the same dataset equally well. To lift these degeneracies, the problem is generally regularised using, for example, Tikhonov regularisation \citep[e.g.][]{Khokhlov1986,piskunov1990}, or the maximum entropy method \citep{vogt1983,skilling1984,vogt1987}. As pointed out by \citet{luger2021}, these methods, albeit well understood \citep[e.g.][]{donati1997a,piskunov2002,kochukhov2002,rice2002}, are both known to alter the reconstructed maps (e.g. with spurious artifacts on smaller spatial scales) and are more sensitive to longitudinal than latitudinal variations, especially for near-equator-on stars. Bayesian implementations of the DI algorithm have been proposed recently and offer a promising route to account for the framework's inherent degeneracies and their effect on the inferred stellar maps \citep[e.g.][]{asensioramos2021,luger2021}.

\subsection{Description of the framework}\label{ssec:framework}

In this study, we use the DI code described in \cite{semel1989,donati1989,brown1991,semel1993,donati1997a,donati2006} and \citet{donati2014}, which is designed to perform Doppler and Zeeman-Doppler analyses of unpolarised and polarised spectra, using maximum-entropy regularisation. This DI framework has been validated on various types of low-mass stars, such as classical and weak-line T Tauri stars \citep[e.g.][]{donati2000,donati2012,donati2014,yu2017,yu2019,finociety2021}, and K/M dwarfs \citep[e.g.][]{hebrard2016,klein2021,zaire2021,zaire2022}.

Our DI framework describes the star's surface as a grid of cells (typically about 10\,000 in this study). Each cell $i$ is assigned a relative brightness coefficient $b_{i}$ (in $[0,1[$ for a dark spot and $>$1 in the case of a bright plage) and a local radial velocity. In the direct approach, the code computes, at all epochs of observation, a local line profile for each grid cell using the Unno-Rachkovsky analytical solution to the radiative transfer equations in a Milne-Eddington atmosphere\footnote{Note that the Unno-Rachkovsky framework is used by default to compute local line profiles in our DI code. Equally-good results would be obtained with more simplistic models like Voigt profiles since (i)~the Sun is slowly rotating and, thus, the intrinsic line profile is unknown (see Sec.~\ref{ssec:221}), (ii)~we are working with cross-correlation functions, whose link with the physical properties of the stellar atmosphere remains obscure, and (iii)~we are only interested in the relative variations of line profiles.} \citep{unno1956,LandiDeglInnocenti2024}. At a given timestamp (i.e. rotational phase), the algorithm computes the sky-projected area $\omega_{i}$ and the RV of each cell of the star. Disk-integrated line profiles are computed by averaging the local profiles, shifted at the local RV, and weighted by the local limb-darkening coefficient, projected cell area and relative brightness. The limb darkening coefficients are computed following the linear law $1-\epsilon(1-\mu)$, where $\epsilon$ is the limb-darkening coefficient \citep[typically taken from][]{claret2013} and $\mu$ is the cosine of the limb angle.

In the inverse approach, our DI framework starts from an unspotted star (i.e. $b_{i} = 1$ for all the cells) and uses a conjugate-gradient process to iteratively compute relative brightness maps until the synthetic line profiles (computed from the model stellar surface) match the observed profiles down to a given level of $\chi^{2}$. The degeneracy of this inversion problem is lifted by imposing a maximum-entropy regularisation condition on our estimator \citep[][]{skilling1984}. In our case, the entropy $\mathcal{S}$ is defined by the Shannon entropy of the relative brightness of the cells:

\begin{equation}
    \mathcal{S} = - \sum_{i}^{N_{\mathrm{c}}} \omega_{i} b_{i} \left[ \log \left( \frac{b_{i}}{b_{0}} \right) -1 \right],
    \label{eq:entropy}
\end{equation}

\noindent
where $N_{\mathrm{c}}$ is the total number of cells at the surface of the star. Conceptually, this can be seen as choosing the brightness distribution with the filling factor f$_{\mathrm{DI}}$ of active regions that can reproduce the observed profiles at a given level of $\chi^{2}$. This filling factor is defined as the sum of the filling factors in dark spots $\left( \sum_{i}\omega_{i}(1-b_{i})/N_{\mathrm{c}} \right)$ and in bright plages  $\left( \sum_{i}\omega_{i}(b_{i}-1)/N_{\mathrm{c}} \right)$.

\subsubsection{Applying Doppler Imaging to slow rotators}\label{ssec:221}

Applying DI to slowly-rotating stars (which we define as having a rotational velocity \vs\,$\lesssim$\,10\,\kms) is more complicated for two main reasons. The first, and most intuitive, is that the angular size of the resolution element at the stellar surface scales as $(\mathcal{R} $\vs$)^{-1}$, where $\mathcal{R}$ is the resolving power of the instrument \citep[see Sec.~9.2.2. of][]{kochukhov2016}. Therefore, in principle, a single snapshot can only probe the largest spatial scales of the brightness variations. This limitation can nonetheless be overcome with a dense temporal sampling of the star with highly-stable high-precision spectrographs, capturing the evolution of the activity-induced distortions as the star rotates and the active regions evolve. In other words, as long as (i)~the SNR of the line profiles is large enough to resolve the activity-induced distortions, and (ii)~enough observations are collected on the time scales over which the star's activity evolves, DI can be, in principle, applied to very slow rotators \citep[e.g., down to \vs\ as low as 1\,\kms\ in][]{hebrard2016}{}.

The other and the most challenging limitation comes from the fact that the intrinsic profile (i.e. the local profile within each DI cell) is unknown. Unlike fast rotators, whose line profiles are generally dominated by the star's rotational broadening, the shape of the disc-integrated line profiles of slow rotators encompasses the effects of formation temperature, micro-turbulence, granulation, magnetic field and rotation, which makes it impossible to predict the intrinsic profile to the level of the photon noise. We must thus assume a local line profile (in the Unno-Rachkovsky’s framework), which will likely exhibit systematic differences with the actual local line profile of slow rotators like the Sun (e.g. because the 3D effects of convection are ignored). In this case, our DI code will naturally focus on reconstructing these systematic differences rather than the activity-induced distortions, leading to unrealistic brightness distributions. To overcome this problem, we use the iterative process introduced in \citet{klein2021} and \citet{klein2022}. We use our DI code to perform a maximum-entropy fit of all the observed CCFs, $I_{\rm{obs}}$. We subtract the median difference between $I_{\rm{obs}}$ and $I_{\rm{syn}}$, the best-fitting synthetic profiles, from $I_{\rm{obs}}$ and repeat the procedure until the median difference between $I_{\rm{obs}}$ and $I_{\rm{syn}}$ is flat (i.e. with a dispersion significantly lower than the photon noise). This process should preserve most of the activity-induced CCF distortions (as confirmed in Sec.~\ref{sec:Results}), but could affect axisymmetric structures generating non-modulated profile distortions though.

\subsubsection{The specific case of Sun-like stars}

Since active regions at the surface of the Sun evolve on a time scale similar to the rotation period \citep[e.g.][]{leighton1964,foukal1998,klein2024b}, the solar disc will likely look dramatically different from one solar rotation to the next. Our DI framework assumes that active regions do not evolve intrinsically, to avoid adding further degeneracies to the model\footnote{Note that recent efforts to account for the evolution of stellar magnetic fields using Gaussian Processes in ZDI reconstructions have shown promising results \citep{finociety2022}.}. In this study, we therefore apply DI on data covering no more than a 27-d rotation cycle, as described in Sec.~\ref{ssec:chunks}.

% this is one of the most serious limitation of DI wrt GPs really - this was mitigated in ZDI by, eg, the TIMES approach of Finociety & Donati 2022, but a similar process is still to be implemented for brightness distributions

Furthermore, our DI code assumes that the profile distortions are only due to brightness inhomogeneities at the stellar surface. This means that the inhibition of the convective blueshift within faculae, which dominates the solar activity RV budget \citep[e.g.][]{meunier2010}, is approximated by brightness contrast effects in our model. The flexibility of DI will naturally reproduce the effects of faculae on the line profiles, even if it means that the resulting brightness distribution loses some of its realism. This will not be a major issue for this study, which focuses on the mitigation of stellar activity signals for RV planet searches. Nonetheless, the realism of the DI maps will be assessed in Sec.~\ref{ssec:4.2}. Additionally, note that our code does not include the effects of granulation and supergranulation \citep[e.g.][]{meunier2015}.

% We therefore split our input data set into 27-d subsets (hereafter called chunks), listed in Table~\ref{tab:list_chunks} and illustrated in Fig.~\ref{fig:chunks}, which we analyse independently in the rest of the paper. We discard all chunks with either fewer than 10 observations or with a gap larger than 25\% in the Sun's rotational phase. These asusmptions, discussed in more details in Section~\baptiste{XXX}, are motivated by the fact that we want to capture activity-induced profile distortions as they are modulated with the Sun's rotation. Thus we need a dense sampling for each rotation cycle that we want to model. Our input data set contains 486 observations spread into 24 chunks, with an average of 20 points per chunk. 

\section{Input data and pre-processing}\label{sec:data}

\subsection{Observations and data processing}

The cross-dispersed échelle spectrograph HARPS-N, at the 3.58-m Telescopio Nazionale Galileo (TNG) at Roque de Los Muchachos observatory (La Palma, Spain), has been monitoring the Sun since 2015 \citep[][Dumusque et al., in prep.]{cosentino2012,dumusque2015,phillips2016,cameron2019,dumusque2021}. Disc-integrated solar spectra spanning most of the optical domain (383 to 690 nm) at a resolving power of $\mathcal{R}$\,=\,115\,000 are collected at a 5-min cadence during day time. Our input data set contains the observations acquired between December 2021 and January 2024 (i.e. in the start of solar cycle 25), reduced with version 3.0.1 of the ESPRESSO data reduction software \citep[DRS;][]{pepe2021,dumusque2021}. Working with these two years of data is motivated by two main reasons. Firstly, the Sun is particularly active and the RV signals exhibit clear quasi-periodic modulations, well suited for DI. Secondly, the data were collected after the refurbishment of the instrument's CCD camera in 2021, and no major maintenance operations were carried out on the instrument since then.

We use the same approach as \citet{klein2024b} to build the input data set. Following \citet{cameron2019}, we only select epochs for which the probability that the Sun is partially masked by clouds in the Earth atmosphere is less than 10\%. Similarly, observations with velocity corrections larger than 0.1\,\ms\ were discarded to ensure that differential extinction is minimum (i.e. that the Sun is close to the zenith). Daily-stacked S1D spectra are continuum-normalised with the open-source software \texttt{RASSINE} \citep{cretignier2020b} and corrected from known instrumental contamination (i.e. gosts, stitchings, interference patterns, ThAr bleeding, defocus of the point spread function and tellurics) using the software \texttt{YARARA} \citep{cretignier2021}.

Cross-correlation functions (CCFs) are computed using the line list tailored for the Sun described in \citet{cretignier2020,cretignier2022}. The RVs, full-width at half-maximum (FWHM) and bisector velocity span \citep[V$_{\mathrm{s}}$;][]{queloz2001} are extracted from the derived CCFs. Our final input data set contains 511 daily-binned CCFs, collected between December 2021 and January 2024 (2.1\,yr) and sampled at the resolution of the instrument \citep[0.82\,\kms;][]{dumusque2021}. The typical photon noise in the CCF continuum is about $2 \times 10^{-5}$, corresponding to a typical RV uncertainty of 0.11\,\ms.

% Since YARARA interpolates each spectrum on a 0.01-\AA\ wavelength grid, the resulting CCFs are slightly oversampled, with velocity bins of 0.53\,\kms\ compared to 0.82\,\kms\ for the instrument \citep{dumusque2021}. Flux uncertainties on the CCFs are artificially boosted by the square-root of the oversampling factor to account for its effect. 

\subsection{Data selection} \label{ssec:chunks}

\begin{figure*}
    \centering
    \includegraphics[width=\linewidth]{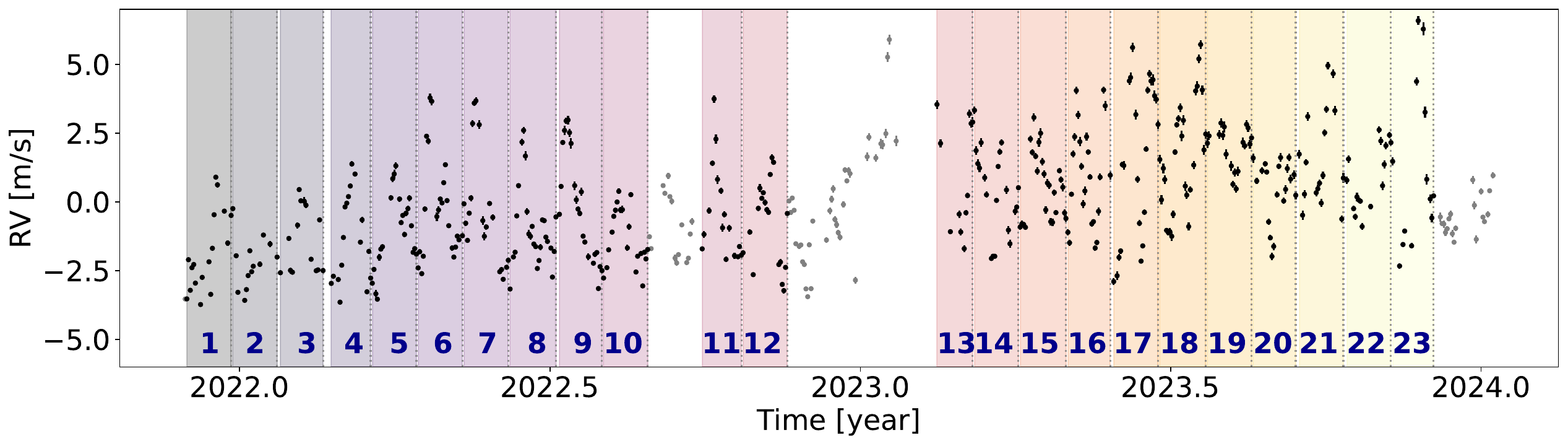}
    \caption{Solar-rest-frame RV time series extracted with HARPS-N DRS. The vertical color bands indicate the different subsets of data used in the DI analysis (see Section 3.2). Grey points were not used in the analysis. The number of each chunk as listed in Tab.~\ref{tab:list_chunks} is indicated in dark blue at the bottom of the figure.}
    \label{fig:chunks}
\end{figure*}

As described in Sec.~\ref{ssec:framework}, we split our input data set into 27-d subsets (hereafter called chunks) that we analyse independently in the rest of the paper. We discard all chunks with either fewer than 10 observations or with a gap larger than 25\% in the Sun's rotational phase. These assumptions, discussed in more detail in Section~\ref{ssec:4.3}, are motivated by the need for a dense temporal sampling, in order to capture the activity-induced profile distortions as they are modulated by the Sun's rotation. Our input data set, shown in Fig.~\ref{fig:chunks}, contains 439 observations spread into 23 chunks, with an average of 19 points per chunk. Information on the different chunks is given in Tab.~\ref{tab:list_chunks}. Finally, the time $t$ of each observation is also expressed in unit of solar rotation phase $\phi_{\mathrm{rot}}$ using 

\begin{equation}
    \phi_{\mathrm{rot}} = \frac{t-T_{0}}{P_{\mathrm{rot}}},
    \label{eq:phase}
\end{equation}

\noindent
where $P_{\mathrm{rot}}$ refers to the Carrington solar rotational period, set to 27.2753 days, and $T_{0}$\,=\,2\,457\,220.55 is a reference time, arbitrarily chosen as the first epoch of the HARPS-N dataset. Note also that this value of $P_{\mathrm{rot}}$ is a good match with the Sun's rotation period measured from the same data set in \citet{klein2024b}.

\begin{table}
\centering
\caption{List of relevant quantities for each of the 23 chunks defined in Section~\ref{ssec:chunks} and shown in Fig.~\ref{fig:chunks}. Columns~2 and~3 give the day and BJD of the epoch corresponding to a rotation phase of zero in Eq.~\ref{eq:phase}, which is also the phase of display of the brightness maps in Fig.~\ref{fig:best_maps}. Columns 4 to 7 give the number of points N$_{\rm{pt}}$, the RV dispersion $\sigma_{\rm{RV}}$, the average continuum dispersion in the CCFs $\sigma_{\rm{CCF}}$, and the recovered active region filling factor} f$_{\mathrm{DI}}$ for each chunk, respectively.
\label{tab:list_chunks}
\begin{tabular}{ccccccc}
\hline
Chunk & Date & BJD & N$_{\rm{pt}}$ & $\sigma_{\rm{RV}}$ & $\sigma_{\rm{CCF}}$ &  f$_{\mathrm{DI}}$  \\
-- & -- & -- & -- & [\ms] & [\%] & [\%] \\
\hline
1 & 2021-12-17 & 59566.23 & 17 & 1.39 & 0.008 & 0.20 \\
2 & 2022-01-14 & 59593.50 & 13 & 0.88 & 0.007 & 0.57 \\
3 & 2022-02-10 & 59620.78 & 14 & 1.13 & 0.007 & 0.36 \\
4 & 2022-03-09 & 59648.05 & 17 & 1.56 & 0.008 & 0.35 \\
5 & 2022-04-05 & 59675.33 & 22 & 1.34 & 0.010 & 0.34 \\
6 & 2022-05-03 & 59702.60 & 23 & 1.80 & 0.010 & 0.26 \\
7 & 2022-05-30 & 59729.88 & 19 & 2.02 & 0.012 & 0.34 \\
8 & 2022-06-26 & 59757.15 & 25 & 1.40 & 0.010 & 0.43 \\
9 & 2022-07-23 & 59784.43 & 22 & 1.93 & 0.012 & 0.37 \\
10 & 2022-08-20 & 59811.70 & 20 & 1.02 & 0.007 & 0.35 \\
11 & 2022-10-13 & 59866.25 & 16 & 1.66 & 0.011 & 0.39 \\
12 & 2022-11-10 & 59893.53 & 19 & 1.47 & 0.007 & 0.29 \\
13 & 2023-02-27 & 60002.63 & 11 & 1.91 & 0.009 & 0.40 \\
14 & 2023-03-26 & 60029.91 & 20 & 1.51 & 0.010 & 0.24 \\
15 & 2023-04-22 & 60057.18 & 25 & 1.19 & 0.006 & 0.29 \\
16 & 2023-05-19 & 60084.46 & 23 & 1.74 & 0.009 & 0.20 \\
17 & 2023-06-16 & 60111.73 & 23 & 2.76 & 0.014 & 0.32 \\
18 & 2023-07-13 & 60139.01 & 27 & 1.98 & 0.009 & 0.22 \\
19 & 2023-08-09 & 60166.28 & 18 & 0.73 & 0.005 & 0.09 \\
20 & 2023-09-06 & 60193.56 & 19 & 1.09 & 0.006 & 0.16 \\
21 & 2023-10-03 & 60220.83 & 17 & 1.68 & 0.008 & 0.31 \\
22 & 2023-10-30 & 60248.11 & 16 & 1.13 & 0.007 & 0.29 \\
23 & 2023-11-26 & 60275.38 & 13 & 2.88 & 0.016 & 0.56 \\
\hline
\end{tabular}
\end{table}

\section{Results}\label{sec:Results}
\subsection{Doppler imaging maps} \label{ssec:Results_DI}

We apply DI to the continuum-normalised HARPS-N Solar CCFs within each 27-d chunk defined in Tab.~\ref{tab:list_chunks}. The inversion process is performed assuming a solar inclination of 80$^{\circ}$ to avoid including additional north-south degeneracies in our fit. This assumption, which is discussed in Sec.~\ref{ssec:4.4}, should not have a major effect on the reconstructed maps since DI is generally not very sensitive to variations of stellar inclinations of 10-20$^{\circ}$, especially for slowly-rotating stars. As a conservative approach, we use CCF continuum dispersion as formal CCF uncertainties, in order to account for sources of noise, such as the instrument instability and stellar variability (e.g. supergranulation), not accounted for in the formal CCF continuum uncertainty. Our final uncertainties are typically around 0.01\%, which is about three times smaller than the RMS of the observed activity-induced distortions in the core of the CCFs. After computing the best-fitting intrinsic profile using the method described in Sec.~\ref{ssec:framework}, the DI fit systematically converges to a reduced $\chi^{2}$ of 1.

\begin{figure*}
    \centering
    \includegraphics[width=\linewidth]{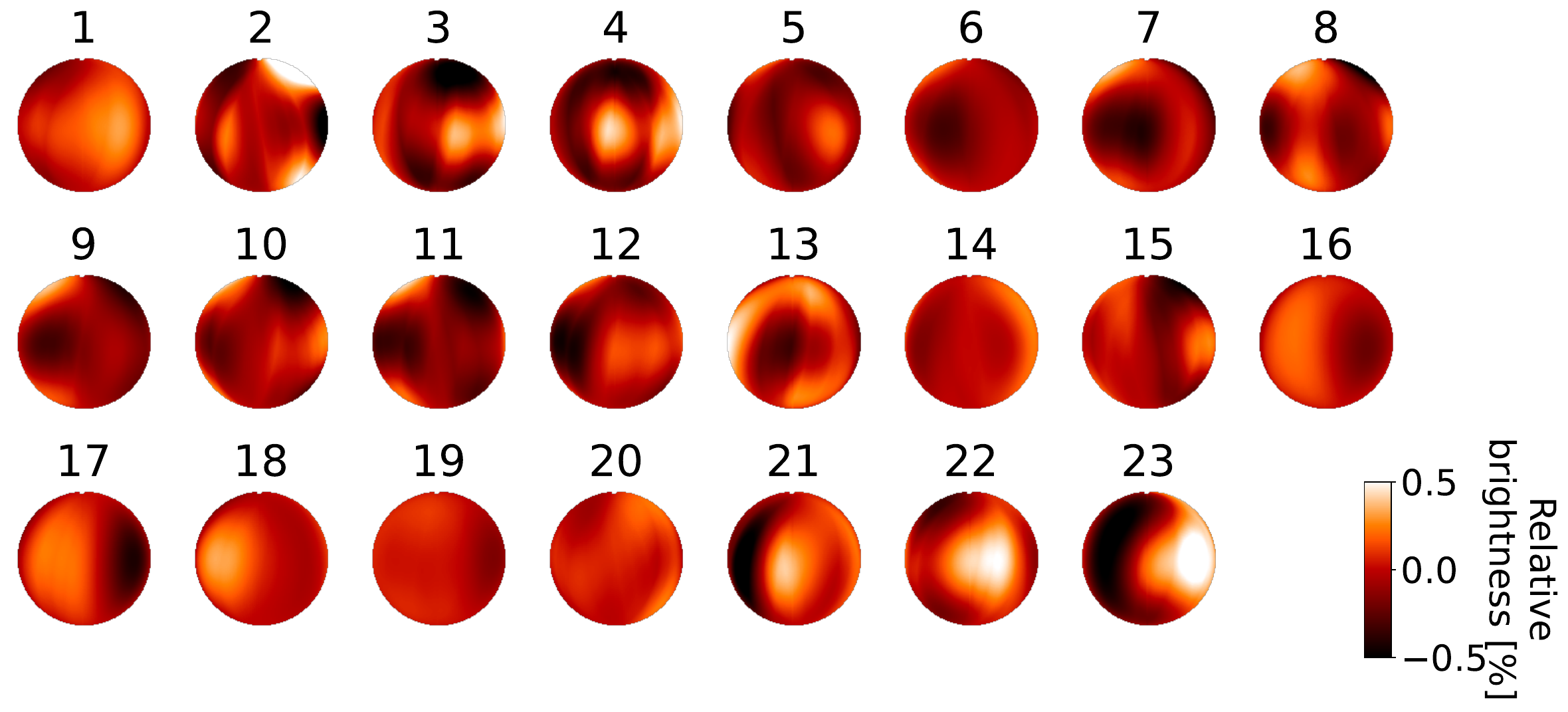}
    \caption{Best-fitting relative brightness distribution of the Sun for each of the 23 chunks listed in Tab.~\ref{tab:list_chunks} and shown in Fig.~\ref{fig:chunks}. The color scale depicts the logarithm of the relative surface brightness. All maps are shown at a solar rotation phase of 0, computed using Eq.~\ref{eq:phase}, and the chunk number is indicated on top of each map.}
    \label{fig:best_maps}
\end{figure*}

\begin{figure*}
    \centering
    \includegraphics[width=1.0\linewidth]{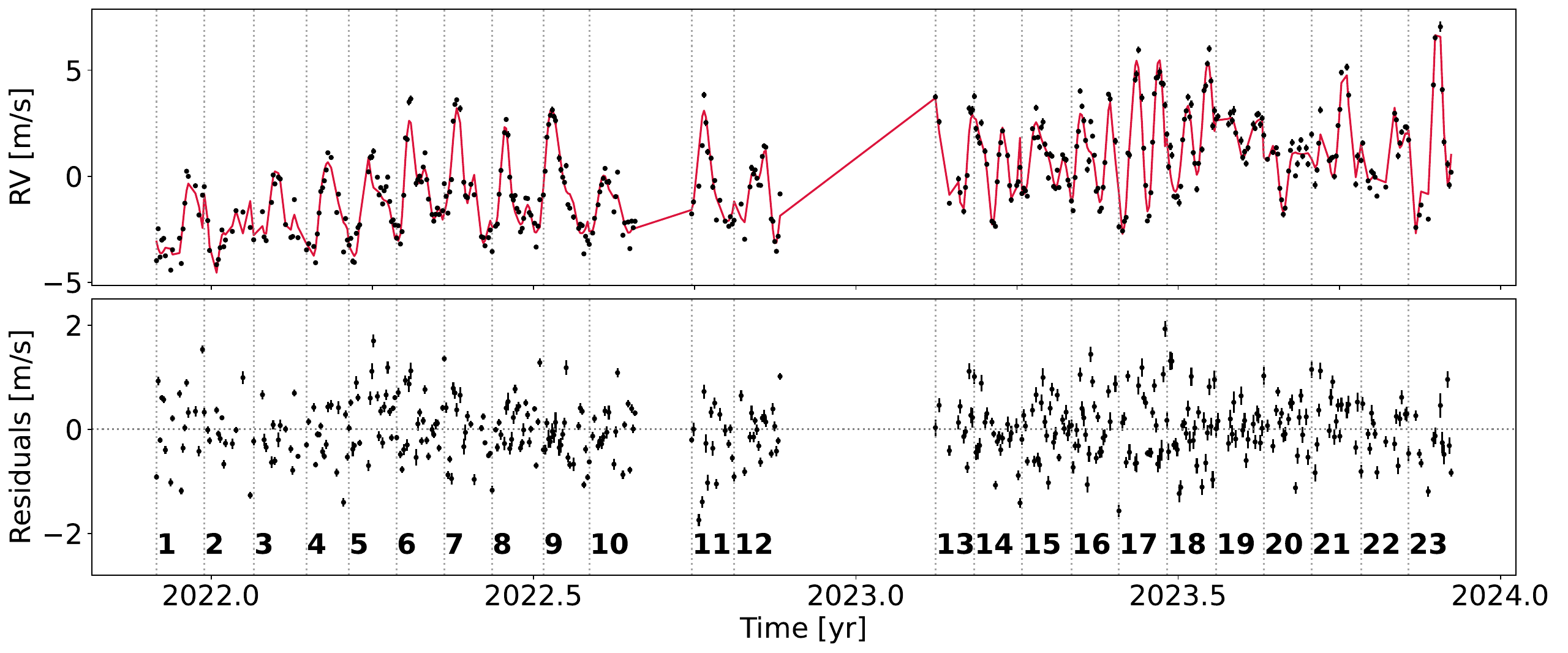}
    \caption{Time series of RVs extracted from HARPS-N DRS (black points) and from the best-fitting DI line profiles (red solid lines, top panel), and residuals (bottom panel). }
    \label{fig:rv_fit}
\end{figure*}

\begin{figure}
    \centering
    \includegraphics[width=\linewidth]{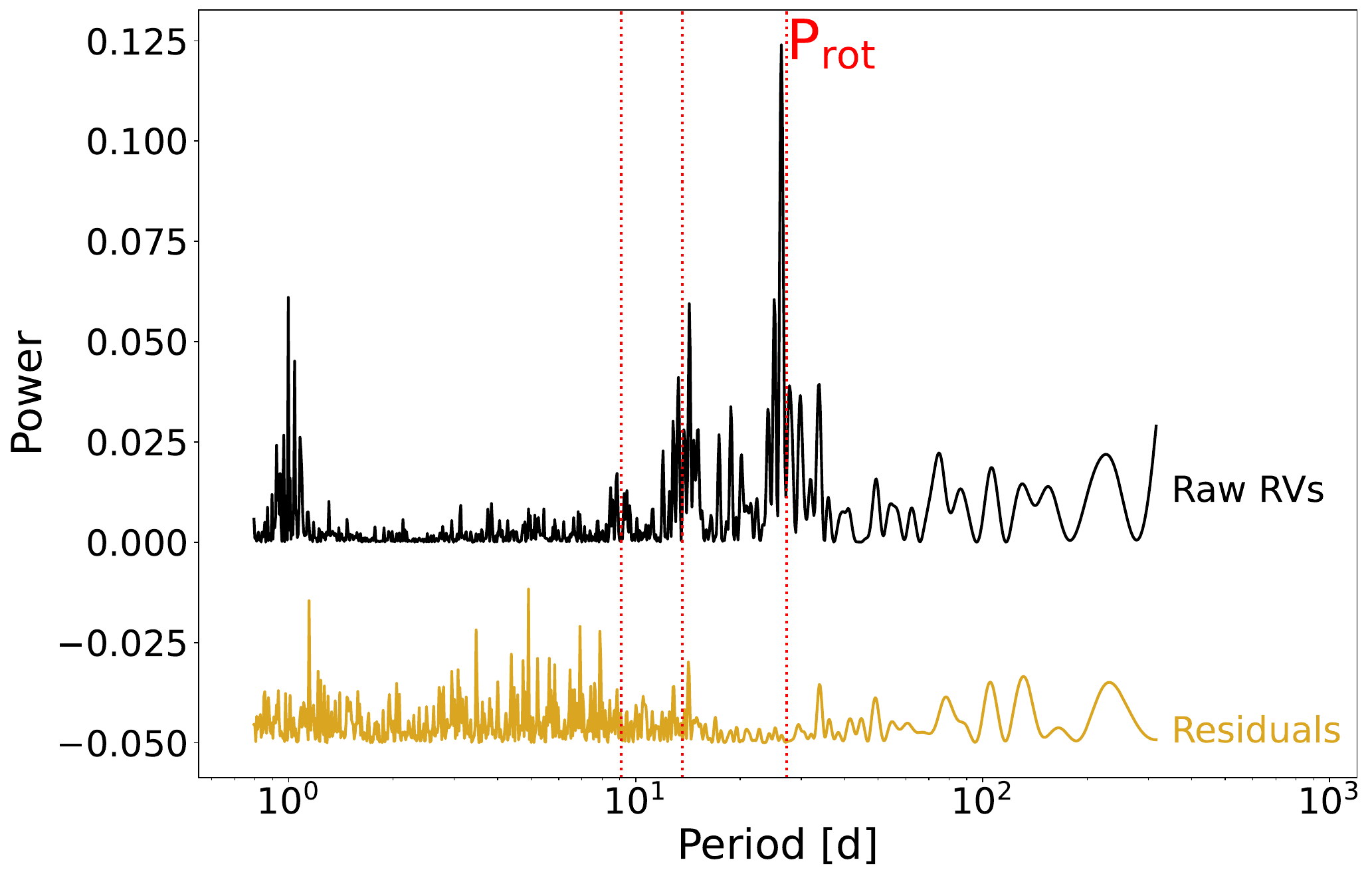}
    \caption{GLS periodogram of the HARPS-N RVs (top black line) and of the RV residuals (bottom yellow line). The false alarm probability of the most prominent peak in the periodogram of the RV residuals, computed using the method of \citet{baluev2008}, is about 0.5. The vertical dotted lines indicate the Sun's rotation period (27.2753\,d here) and first two harmonics.} The periodograms were computed using the \texttt{astropy} python module \citep{astropy2013,astropy2018,astropy2022}.
    \label{fig:rv_per}
\end{figure}

The best-fitting brightness maps, shown in orthographic projection in Fig.~\ref{fig:best_maps}, exhibit filling factors ranging from 0.09\% to 0.57\% (see Fig.~\ref{fig:rect_maps} for the relative brightness distribution on the full solar surface). We also show the best fit to the line profiles in Fig.~\ref{fig:best_profiles_1} and Fig.~\ref{fig:best_profiles_2}. To estimate the typical variation of the filling factor in active regions induced by noise, we create series of mock data sets by adding Gaussian white noise, drawn from the uncertainties adopted on the CCFs (see Tab.~\ref{tab:list_chunks}), to the best-fitting profiles of our model. We then apply DI to the mock CCF time series. For each chunk, we repeat the process for 10 white noise realisations and found a typical relative variation of 20\% on the filling factor. The robustness of the reconstructed maps is discussed in more detailed in Sec.~\ref{ssec:4.3}.

As a sanity check, we extract the RVs from the best-fitting CCFs and compare them to the solar RVs extracted from HARPS-N DRS in Fig.~\ref{fig:rv_fit}. We find that both RV time series match well, with a residuals Root Mean Square (RMS) of 0.58\,\ms, compared to a dispersion of 2.2\,\ms\ in the input RVs. Using generalised Lomb-Scargle (GLS) periodograms \citep{zechmeister2009}, we found no periodicity in the RV residuals (see Fig.~\ref{fig:rv_per}). In particular, most power at the Sun's rotation period and first harmonic has been removed in the residuals, which indicates that the rotationally-modulated activity component is well captured by our model. The RMS of the RV residuals remains significantly larger than the formal RV uncertainties ($\sim$0.1\,\ms) but is consistent with that obtained with a Gaussian process regression on the RV time series \citep[see][]{klein2024b}. Since oscillation- and granulation-induced RV variations have been, for the most part, averaged out in our daily-binned data \citep{dumusque2011,chaplin2019}, the dispersion budget in the RV residuals is most likely a mix of instrument stability \citep[$\sim$0.5\,\ms;][]{dumusque2021} and supergranulation signals \citep[estimated at around 0.7\,\ms, but likely reduced by the daily binning process;][]{meunier2015,almoulla2023,lakeland2024}.

\subsection{Comparison with SDO observations}\label{ssec:4.2}

A clear advantage of using Sun-as-a-star observations is that our best-fit brightness maps of Fig.~\ref{fig:best_maps} can be compared with resolved images of the Sun. The Helioseismic and Magnetic Imager \citep[HMI;][]{schou2012,scherrer2012} of the Solar Dynamics Observatory \citep[SDO;][]{pesnell2012} have been collecting high-cadence high-resolution observations of the solar surface since 2010. We used the open-source pipeline \texttt{SOLASTER} \citep{ervin2022} to download the HMI intensitygrams, dopplergrams and magnetograms at the epochs corresponding to a rotational phase of zero as defined in Eq.~\ref{eq:phase} (i.e. the same phase as the maps shown in Fig.~\ref{fig:best_maps}), for each of the 23 chunks. The Dopplergrams are corrected for the Sun's rotational velocity and spacecraft motion, and intensitygrams are corrected for limb darkening. To make sure that the SDO data are not affected by short term variability (e.g. p-mode oscillations and granulation), we average ten SDO maps collected on a 10-h window centered on the Sun's rotation phase of interest.

We model the SDO dopplergrams, magnetograms and mean-subtracted intensitygrams using spherical harmonics (SH) of maximum degree $\ell_{\mathrm{max}}$\,=\,5, corresponding to the typical spatial resolution probed by our DI algorithm. Hereafter, we refer to the SH model of the SDO maps as large-scale SDO maps. As a starting point, we compare the orthographic projection of our DI brightness maps at rotational phase 0 (as shown in Fig.~\ref{fig:best_maps}, to the large-scale SDO outputs at the same phase. We find that the Pearson correlation coefficient between the large-scale SDO dopplergrams and the DI brightness maps varies significantly from one chunk to the next, with values as high as 0.76, but with mean absolute value of 0.34 and standard deviation of 0.22. Correlation coefficients with the large-scale SDO intensitygrams and magnetograms do not reach absolute values greater than 0.5.

\begin{figure}
    \centering
    \includegraphics[width=\linewidth]{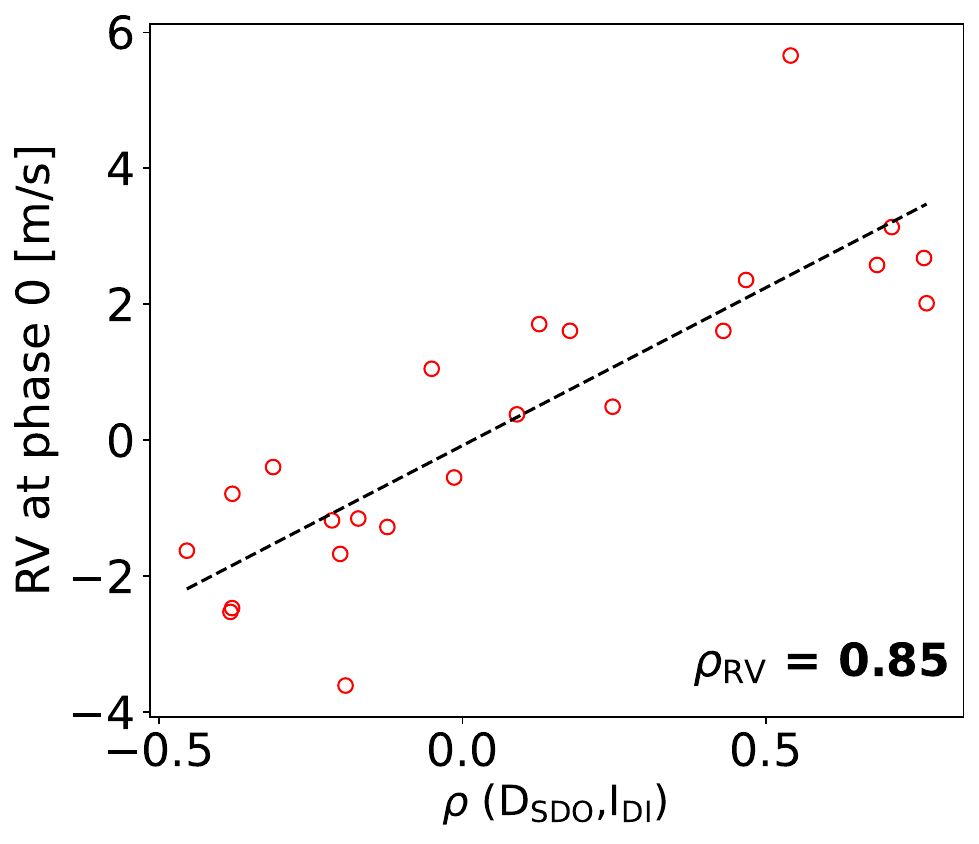}
    \caption{Correlation coefficients between SDO large-scale dopplergrams (D$_{\mathrm{SDO}}$) and DI brightness maps (I$_{\mathrm{DI}}$), against the solar RVs at rotational phase 0, as defined in Eq.~\ref{eq:phase}. The two time series exhibit a Pearson correlation coefficient $\rho_{\mathrm{RV}}$ of 0.85.}
    \label{fig:correl_rv}
\end{figure}

\begin{figure}
    \centering
    \includegraphics[width=\linewidth]{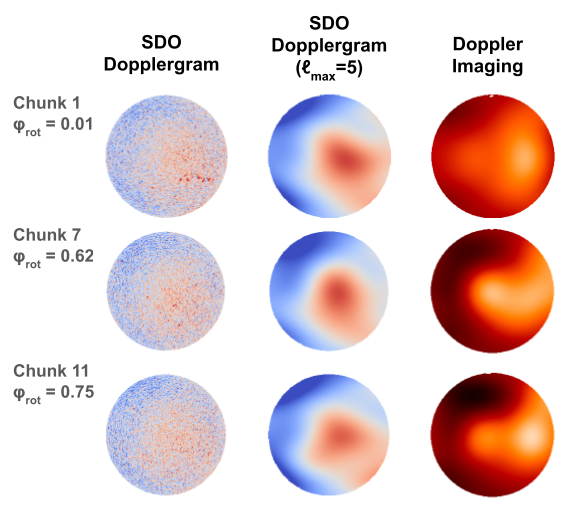}
    \caption{Qualitative comparison between SDO dopplergrams and DI brightness maps at the rotational phase that maximises the absolute RV value within chunk 1, 7 and 11, as defined in Tab.~\ref{tab:list_chunks}. From left to right, we show the SDO dopplergram corrected from rotational flows, the spherical harmonic model of the SDO dopplergram with $\ell_{\mathrm{max}}$\,=\,5, and the best-fit relative brightness maps obtained with DI. The color bars are defined between $\pm$0.5\kms, $\pm$0.15\kms, and $\pm$0.5\%, for the left-, middle- and right-hand column, respectively.}
    \label{fig:compar_sdo}
\end{figure}

As shown in Fig.~\ref{fig:correl_rv}, the correlation coefficients $\rho$ between the DI maps and SDO dopplergrams are themselves correlated with the RV (resp. FWHM) at phase 0, with a Pearson correlation coefficient of 0.85 (resp. 0.75). We interpret this as follows. Line profile distortions inducing RV signals of absolute value smaller than $\sim$2\,\ms\ have an amplitude comparable to the dispersion of the CCF continuum and, therefore, are not well separated from the noise in the DI inversion. On the other hand, the positions of active regions inducing RV signatures with absolute values larger than $\sim$2\,\ms\ are accurately recovered within $\sim$36$^{\circ}$ (i.e. the typical angular resolution of SH with $\ell_{\mathrm{max}}$\,=\,5). To bring further evidence to this suggestion, we compared the large-scale SDO dopplergrams to the DI maps, this time at the stellar phase which maximises the absolute value of the RV in each chunk (see Fig.~\ref{fig:compar_sdo}). This time, the median absolute value of $\rho$ rises to 0.63\,$\pm$\,0.1, with a maximum of 0.83 (see Fig.~\ref{fig:all_corr_coef}). We also note that positive RVs (resp. negative RVs) lead to positive (resp. negative) correlation coefficients between large-scale SDO dopplergrams and DI maps. Since the input line profiles of each chunk are centered (i.e. the RVs are mean-subtracted), negative RVs correspond to negative correlations between the DI maps and SDO dopplergrams.

The fact that $\rho$ does not reach greater values than 0.83 and varies significantly from one rotation to the next, even at the phase of maximum distortion, can be explained by several factors. Firstly, the accuracy of the DI maps is intrinsically limited by the fact that the mean absolute value of the maximum RV of each chunk is only 1.8\,\ms. Secondly, our assumed stellar inclination of 80$^{\circ}$ will slightly affect the latitude of the active regions. In addition, choosing the maximum entropy maps will likely favour maps with more equatorial active regions. These two factors can induce mismatches with the solar surface, especially at the beginning of the magnetic cycle, where active regions appear at higher latitudes. Finally, the presence of more complex signals on the SDO dopplergrams (e.g. supergranulation), not included in our DI framework, could also affect the comparison with the brightness maps.

We also find a significantly better match between the large-scale SDO intensitygrams and the DI maps at the phase of maximum profile distortion. The absolute value of the correlation coefficient between these maps increases from 0.11\,$\pm$\,0.02 to 0.32\,$\pm$\,0.10, with a maximum of 0.7. Since the activity-induced distortions in the line profiles are dominated by the inhibition of convective blueshift in faculae \citep[e.g.][]{meunier2010,meunier2015,milbourne2019,meunier2022}, we naturally expect the DI maps to be better matched by the SDO dopplergrams than the intensitygrams, which are primarily sensitive to sunspots. At the stellar phase of maximum profile distortion, the locations of the most prominent regions of convective blueshift inhibition are relatively well recovered in our brightness maps. Since faculae tend to cluster around sunspots, the brightness maps will naturally roughly recover the position of sunspots, hence the increased correlation with the large-scale SDO intensitigrams. Conversely, we do not observe good matches between the DI maps and the large-scale SDO magnetograms (or their absolute values), probably because of the higher level of complexity of the small-scale magnetic field network.

\subsection{Robustness tests}\label{ssec:4.3}

As shown in the previous section, Doppler imaging is able, in most cases, to accurately reproduce large-scale SDO dopplergrams from densely-sampled HARPS-N solar spectra. This naturally raises the question of its application to other stars. In this section, we perform simulations to estimate the accuracy of DI maps as a function of the number of points, noise level and RV value. From one representative DI maps (chunk~7 in this case), we generate time series of synthetic profiles, to which we add normally-distributed noise. The synthetic profiles are then modelled with DI and the best-fitting map is compared to the input brightness distribution at different stellar phases (i.e. RV values). In each case we select a number N$_{\mathrm{pt}}$ of epochs (between 5 and 20) and randomly assign their stellar phases between 0 and 1\footnote{Note that our assumption to limit our observations to one stellar rotation phase is purely driven by the fact that, for the Sun, active regions evolve on the time scale of one rotation period. For other stars, chunks should be defined on the time scale on which stellar activity varies rather than on one rotation phase.}. Relative noise levels between $10^{-5}$ and $3 \times 10^{-4}$ are considered, the latter being of similar amplitude to the largest distortions in the input synthetic profiles. We repeat the simulation ten times for each number of epochs and noise level, each time randomly assigning different observational phases and noise realisation to the synthetic profiles.

\begin{figure}
    \centering
    \includegraphics[width=\linewidth]{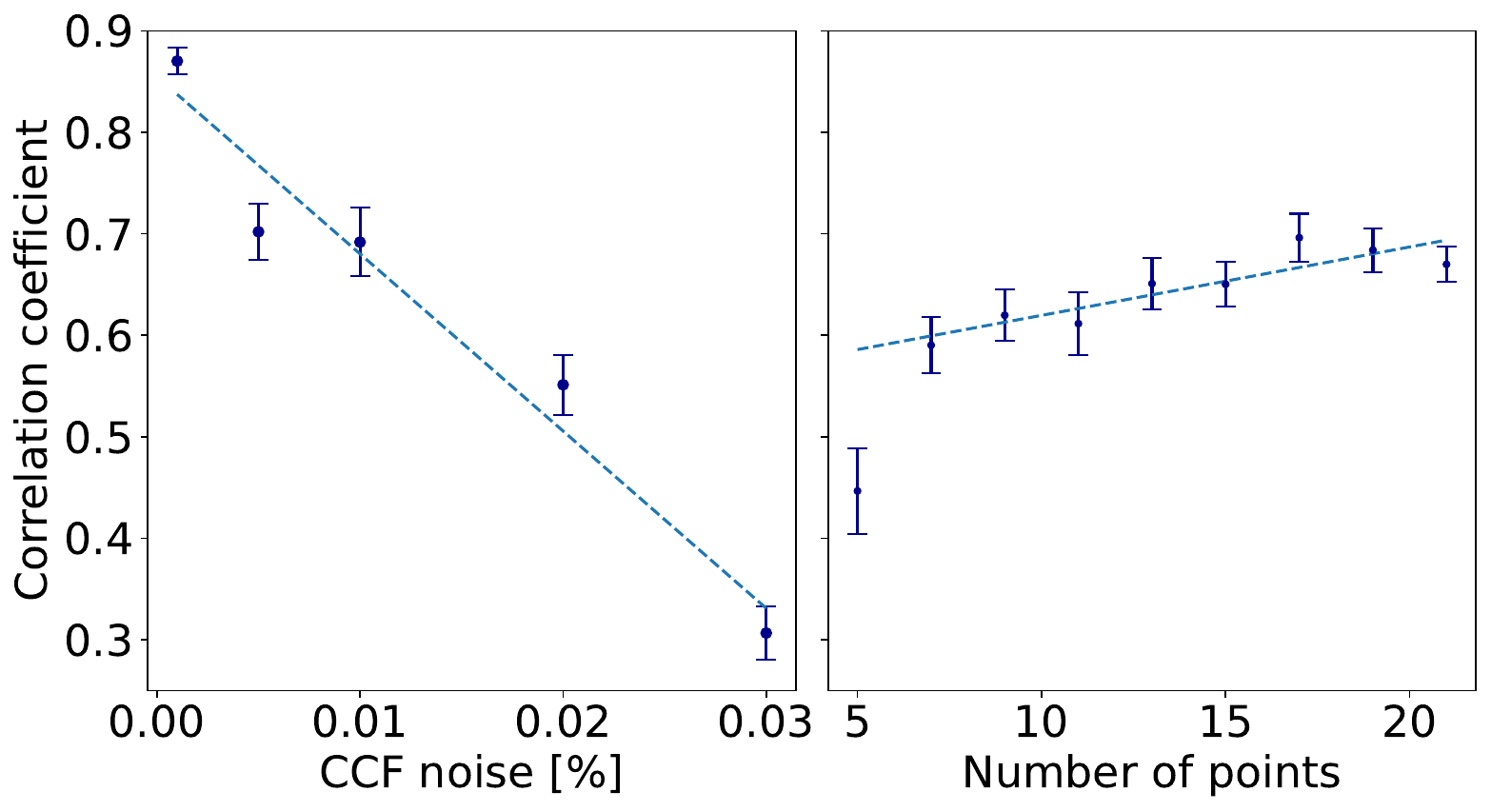}
    \caption{Average Pearson correlation coefficient between the input brightness distribution and the recovered DI maps as a function of the noise level in the continuum-normalised line profiles (left panel), and the number of epochs (right panels). In both panels, the dashed lines indicate the best-fiting straight line (note that, on the right panel, the first point is not included in the linear fit).}
    \label{fig:syn_noise}
\end{figure}

Fig.~\ref{fig:syn_noise} shows the evolution of the Pearson correlation coefficient between the input and recovered brightness maps, as a function of the CCF noise per observation and number of observations. We find that the correlation coefficient between the input and the recovered brightness distributions linearly increases with the number of epochs for N$_{\mathrm{pt}}$\,$\gtrsim$\,7. In most cases, a smaller number of observations per rotational phase does not ensure a sufficient phase coverage to accurately recover the input brightness distribution. The correlation coefficient decreases roughly linearly with the noise level, with values consistently larger than 0.5 for relative levels larger or equal to $\sim$10$^{-4}$ with respect to the continuum, corresponding to the typical amplitude of the activity-induced distortions in the input data. It is worth noting that, for more than 7 observations, it is the CCF noise rather than the number of points that controls the accuracy of the retrieved brightness map. As expected, we also find that the input map is more accurately reproduced at phases with larger RV values. A two-dimensional color map of the correlation between the input and retrieved brightness maps as a function of the number of observations and CCF noise is provided in Fig.~\ref{fig:correl_map_noise}.

\subsection{Impact of stellar parameters}\label{ssec:4.4}

\subsubsection{Differential rotation}

Doppler imaging is a powerful tool to constrain the latitudinal differential rotation (DR) of rapidly-rotating stars \citep[e.g.][]{donati2000,petit2002,yu2019,zaire2021}. Assuming a solar-like DR, the rotation rate $\Omega$ as a function of the colatitude $\theta$ is given by

\begin{equation}
    \Omega (\theta) = \Omega_{\mathrm{eq}} - \left( \cos \theta \right)^{2} \mathrm{d} \Omega ,
    \label{eq:DR}
\end{equation}

\noindent
where $\Omega_{\mathrm{eq}}$ is the rotation rate at the stellar equator and d$\Omega$ is the difference in rotation rate between the equator and the pole. Traditionally, DI inversions are performed for a grid of ($\Omega_{\mathrm{eq}}$,d$\Omega$) to a fixed level of information (i.e. to a fixed level of $f_{\mathrm{DI}}$), and the best fitting DR parameters are estimated from the resulting $\chi^{2}$ map. This method will most likely fail in the case of the Sun, since active regions evolve on time scales similar to the rotation period. Yet, our input data set covers several consecutive solar rotation cycles. Thus, by studying how the position of the largest features evolves between consecutive maps, one may still be able to place constraints on the solar rotation and DR.

The optimal brightness distribution of each chunk is projected into an rectangular view, and normalised by its standard deviation. For each pair of consecutive rectangular projections, we differentially rotate the rows of the first map for a grid of ($\Omega_{\mathrm{eq}}$,$\mathrm{d} \Omega$) values using Eq.~\ref{eq:DR}, and compute the correlation between the rotated map and the second map. The inferred DR parameters are obtained by modelling the correlation map with a second-order polynomial near the maximum correlation. Uncertainties on the DR parameters are obtained independently for each pair of consecutive chunks using a bootstrap. This is done by adding normally-distributed Gaussian white noise to the best-fit line profiles of Sec.~\ref{ssec:Results_DI}, and, then, inverting these line profiles into brightness distributions with DI. The best-fitting DR parameters are then computed between the pair of synthetic brightness maps using the technique described above. We repeat this process 200 times with different white noise realisations, and take the standard deviation on the distribution of $\Omega_{\mathrm{eq}}$ and $\mathrm{d} \Omega$ as 1$\sigma$ uncertainties for the parameters. For each chunk, we consider that the Sun's rotation has been detected if the best-fitting value of $\Omega_{\mathrm{eq}}$ lies betweens 0.224 and 0.256\,rad\,d$^{-1}$ (i.e. rotation periods of 28 and 24.5\,d, respectively). We could not draw a direct link between the variation of 
$\Omega_{\mathrm{eq}}$ and the Sun's magnetic cycle, which is not surprising as the timescale of our observations is only two years.

\begin{figure}
    \centering
    \includegraphics[width=\linewidth]{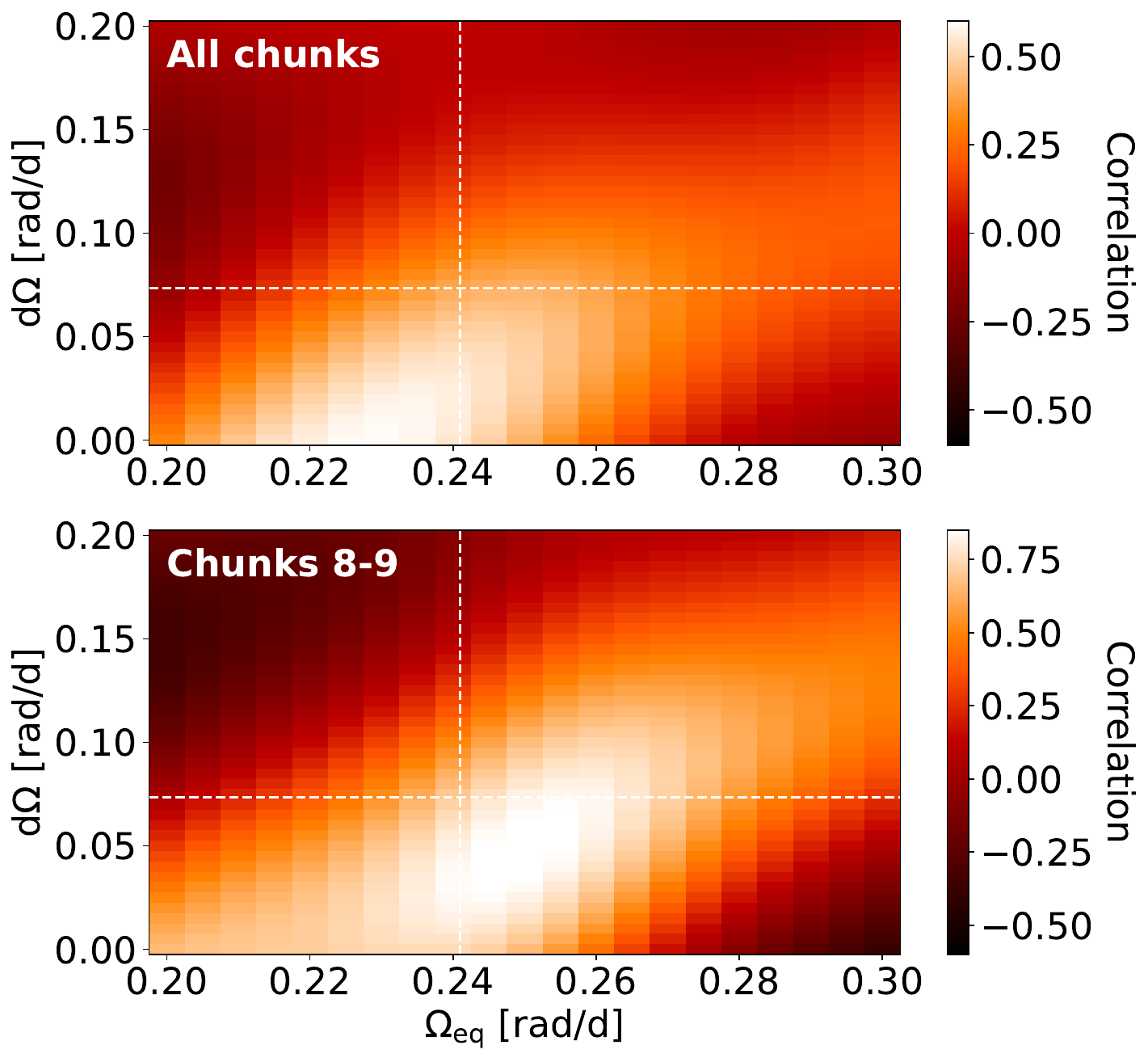}
    \caption{Correlation between consecutive DI brightness maps averaged over all pairs of chunks with detected solar rotation (top panel) and between chunks 8 and 9 in Tab.~\ref{tab:list_chunks} (bottom panel), in the ($\Omega_{\mathrm{eq}}$,$\mathrm{d} \Omega$) space. In both panels the white dashed lines indicate the solar DR values adopted in the study.}
    \label{fig:diff_rot}
\end{figure}

The top panel of Fig.~\ref{fig:diff_rot} shows the average correlation map for all pairs of chunks with detected solar rotation, in the ($\Omega_{\mathrm{eq}}$,$\mathrm{d} \Omega$) space. The Sun's equatorial rotation period is accurately recovered for 14 of the 20 pairs of brightness maps analysed, with an average equatorial rotation rate of 0.227\,$\pm$\,0.018\,rad\,d$^{-1}$ (i.e. a rotation period of 27.7\,$\pm$\,2.1\,d at the equator). This value is fully consistent with the solar rotation period measured by \citet{klein2024b} from the same dataset with Gaussian Processes (27.4\,$\pm$\,0.4\,d). In contrast, the solar rotation rate is not well retrieved when at least one of the chunks exhibit a low activity level (i.e. a peak-to-peak RV variation smaller than $\sim$3\,\ms, like in Chunks 2 and 19). We also report a marginal 2.5$\sigma$ detection of DR between chunks 8 and 9, with $\mathrm{d} \Omega$\,=0.05\,$\pm$\,0.02 rad\,d$^{-1}$, consistent with the solar value of $\sim$0.07\,rad\,d$^{-1}$ (see the bottom panel of Fig.~\ref{fig:diff_rot}). However, the Sun's differential rotation remains undetected in all other pairs of chunks, which is most likely due to the poor latitudinal resolution of our brightness maps, and the fast-evolving solar activity.

\subsubsection{Stellar Inclination}

It is challenging to constrain the inclination of the star's rotation axis from the activity-induced distortions of unpolarised line profiles, especially for slowly-rotating stars. We found that, at the level on one chunk, near-pole-on configurations (i.e. inclinations smaller than $\sim$30$^{\circ}$) can be confidently excluded, but that no precise estimate of the star's inclination could be recovered, given the relatively small length of the chunks and angular resolution of the recovered brightness maps.

In an attempt to refine the constraints on the star's inclination, we performed a DI inversion of all the 439 line profiles simultaneously, whilst varying the assumed stellar inclination in the fit. The resulting grid of reduced $\chi^{2}$ reaches a minimum of 2.2, significantly larger than 1, due to the fact that activity evolution is not accounted for in the model. However, we find a minimum in $\chi^{2}$ grid, yielding a stellar inclination estimate of 69\,$\pm$16$^{\circ}$. Inclinations smaller than $\sim$40$^{\circ}$ are rejected at a 3$\sigma$ level. We conclude that DI can help setting lower limits on stellar inclinations, but cannot provide precise constraints on this parameter for near-equator-on stars. In particular, our assumption of an inclination of 80$^{\circ}$ in the DI process does not have a major impact on the results of this paper.

\section{Planet injection and recovery} \label{sec:planets}

We perform a series of planet injection-recovery tests in order to estimate the sensitivity of the DI framework to low-mass planet signatures. We consider the three different planet signatures listed in Tab.~\ref{tab:planet_prop}. Two short-period planets with an orbital period $P_{\mathrm{orb}}$ of 6\,d and RV semi-amplitudes of 1\,\ms\ (our fiducial case) and 0.4\,\ms\ (Earth-mass planet), and one super-Earth at an orbital period of 100\,d. In each case, we linearly interpolate the observed CCFs and shift them according to the RV signature of a single of the three planets. The planet orbit is assumed circular, and, thus, the RV signature $v_{\mathrm{p}}$ as a function of time $t$ is given by

\begin{equation}
    v_{\mathrm{p}} = K_{\mathrm{p}} \sin 2 \pi \left[  \frac{T_{0}-t}{P_{\mathrm{orb}}} + \phi_{\mathrm{p}} \right],
    \label{eq:planet}
\end{equation}

\noindent
where $K_{\mathrm{p}}$, $P_{\mathrm{orb}}$ and $\phi_{\mathrm{p}}$ are the planet RV semi-amplitude, orbital period and orbital phase, respectively, and where T$_{0}$ is the reference time of Eq.~\ref{eq:phase}. For each planet, we consider three different orbital phase to ensure that the results are not biased. Note that we also extract the CCF RVs, $v_{\mathrm{obs}}$, using a Gaussian fit. Multi-planetary systems are not considered in this proof-of-concept study.

\begin{table}
    \centering
    \caption{Orbital periods P$_{\mathrm{orb}}$, RV semi-amplitudes K$_{\mathrm{p}}$, masses M$_{\mathrm{p}}$ and orbital phases $\phi_{\mathrm{p}}$ of the synthetic planets considered in this study.}
    \label{tab:planet_prop}
    \begin{tabular}{cccccc}
    \hline
      \textbf{Case}  & P$_{\mathrm{orb}}$ & K$_{\mathrm{p}}$ & M$_{\mathrm{p}}$ & $\phi_{\mathrm{p}}$  \\
      -- & [d] & [\ms] & [M$_{\oplus}$] & -- \\
      \hline
      1 & 6.0 & 1.0 & 2.8 & \{0.0, 0.33, 0.7\} \\
      2 & 6.0 & 0.4 & 1.1 & \{0.0, 0.33, 0.7\} \\
      3 & 100.0 & 0.5 & 3.7 & \{0.0, 0.33, 0.7\} \\
      \hline
    \end{tabular}
\end{table}

\begin{table}
    \centering
    \caption{Best estimates of the orbital period, RV semi-amplitude and orbital phase of the planet signatures described in Tab.~\ref{tab:planet_prop}, when retrieved from the DI-corrected RVs.}
    \label{tab:D1_results}
    \begin{tabular}{ccccc}
    \hline
        Case & P$_{\mathrm{orb}}$ & K$_{\mathrm{p}}$ & $\phi_{\mathrm{p}}$  \\
        -- & [d] & [\ms] & -- \\
    \hline
     &  6.002\,$\pm$\,0.002 & 0.58\,$\pm$\,0.04 & 0.02\,$\pm$\,0.02 \\
 Planet 1  & 6.000\,$\pm$\,0.002 & 0.57$^{+0.04}_{-0.05}$ & 0.33$\pm$0.02 \\
    & 6.000\,$\pm$\,0.002 & 0.62\,$\pm$\,0.04 & 0.68$\pm$0.02 \\
    \hline
    & 6.006\,$\pm$\,0.005 & 0.23$^{+0.04}_{-0.05}$ & 0.04$^{+0.07}_{-0.06}$ \\
  Planet  2  & 5.997\,$\pm$\,0.005 & 0.24$^{+0.05}_{-0.06}$ & 0.31$^{+0.06}_{-0.04}$ \\
       & 5.998\,$\pm$\,0.004  & 0.29$^{+0.04}_{-0.05}$ & 0.66$\pm$0.05 \\
    \hline
    & 99.5$^{+0.8}_{-1.0}$ & 0.35$^{+0.05}_{-0.06}$ & -0.01$\pm$0.04 \\
   Planet   3 & 100.2$^{+1.0}_{-1.2}$ & 0.29$^{+0.04}_{-0.05}$ & 0.33$\pm$0.04 \\
       & 99.5$^{+0.7}_{-0.8}$ & 0.29$^{+0.04}_{-0.05}$ & 0.68$^{+0.04}_{-0.03}$ \\ 
    \hline
    \end{tabular}
\end{table}

\subsection{Doppler Imaging as a planet detection tool}\label{ssec:5.1}

We first perform a simple test to assess the ability of DI to be used as an activity filtering tool. As in Section~\ref{ssec:Results_DI}, we blindly model the CCFs using DI, regardless of the presence of planet signatures in the data. From the best-fitting brightness distributions, we generate synthetic CCFs at the epochs of observations, compute the RVs of the synthetic CCFs, and subtract these RVs from $v_{\mathrm{obs}}$. Ideally, the DI correction should remove most of the stellar activity signals whilst preserving the injected planet signature.

\begin{figure}
    \centering
    \includegraphics[width=\linewidth]{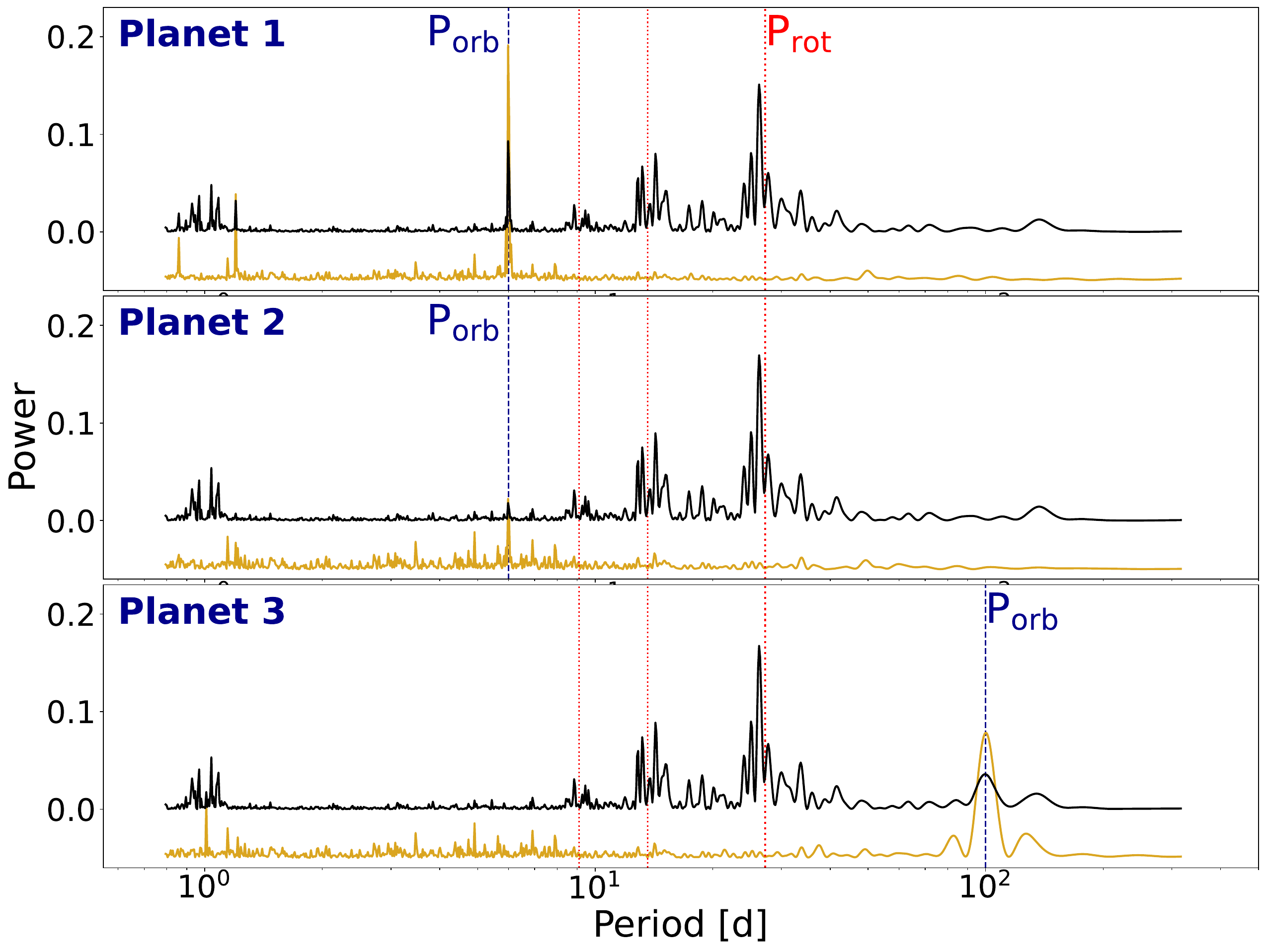}
    \caption{GLS periodograms of the HARPS-N solar RVs after the injection of the planet RV signatures listed in Tab.~\ref{tab:planet_prop}. In each case, the top black and bottom yellow lines represent the periodogram of the RV time series before and after applying DI to filter stellar activity signals. In each panel, the blue vertical dashed line indicates the orbital period of the injected planet, and the three red dotted lines mark the Sun's rotation period and first two harmonics.}
    \label{fig:D1_periodograms}
\end{figure}

The GLS periodograms of the DI-corrected RV time-series are shown in Fig.~\ref{fig:D1_periodograms}. Similarly to Section~\ref{ssec:Results_DI}, most of the power at P$_{\mathrm{rot}}$ and its harmonics has disappeared in the DI-corrected RVs, which indicates that DI has efficiently filtered the quasi-periodic activity signals from the data. We also note that the most prominent peak in the periodogram of the DI-corrected RVs corresponds, in all cases, to the orbital period of the injected planet. We then model the DI-corrected RVs using Eq.~\ref{eq:planet}, fitting for K$_{\mathrm{p}}$, P$_{\mathrm{orb}}$ and $\phi_{\mathrm{p}}$, as well as for an uncorrelated jitter term, quadratically added to the formal RV uncertainties. The parameter space is sampled using the Bayesian Markov Chain Monte Carlo (MCMC) process implemented in the open-source software \texttt{pyaneti} \citep{barragan2019,barragan2022A}. Non-informative uniform prior distributions are adopted for all the model parameters.

The optimal planet parameters are shown in Tab.~\ref{tab:D1_results}. As intuited from Fig.~\ref{fig:D1_periodograms}, the orbital period and phase of the injected planets are fully consistent with their injected counterparts. However, the recovered RV semi-amplitudes are systematically under-estimated by about 40\%, which suggests that the DI process has partly filtered out the planet signatures. In other words, planet signatures in the CCF can be, to an extent, modelled with DI alone. The planet CCF signature is a simple Doppler shift and, as such, affects only the first derivative of the CCF with respect to the wavelength. On the other hand, stellar activity affects mostly higher-order derivatives, which is the basis of data-driven methods to filter stellar activity in the wavelength-space \citep[e.g.][]{jones2017,cameron2021,klein2024b}. However, there is no reason for the first derivative of the CCF to be orthogonal to higher odd-order derivatives (e.g. due to cross-talks between these derivates and the first-order derivative). This cross-talk could explain why our activity model is able to filter out part of the planet signature. Therefore, similarly to other wavelength-based activity modelling frameworks \citep[e.g.][]{john2022,wilson2022,john2023}, we need to simultaneously model stellar activity signals and planet signatures. However, it is worth noting that the blind application of DI was able to provide accurate estimates for the orbital period and phase of the planets, which can be used as informative priors to other models.

\subsubsection{Planet detection limit}\label{sssec:inj_rec}

Building on the results obtained above, we now assess the ability of DI to be used as a blind planet search tool by generating 1,000 data sets from the solar HARPS-N observations. For each data set, we inject a single planetary signal directly into the same 439 CCFs selected in Sec.~\ref{ssec:chunks} (corresponding to the 23 chunks listed in Tab.~\ref{tab:list_chunks}). The planet signatures are computed using Eq.~\ref{eq:planet}, assuming circular planetary orbits. The planet orbital period, $P_{\mathrm{orb}}$, is randomly drawn from a log-uniform law between 5 and 300 days, whereas $K_{\mathrm{p}}$ and $\phi_{\mathrm{p}}$ are randomly drawn from uniform laws between 0.2 and 1.0\,\ms\ and 0.0 and 1.0, respectively (baseline of 2.1 year).

For each data set, we use three independent methods to filter stellar activity signals. Firstly, we simply model the stellar activity RV signal by a sine-wave at the rotation period of the Sun and its first harmonic. Secondly, we apply the Doppler-constrained principal component analysis (DCPCA) framework introduced in \citet{jones2017}, as implemented in \citet{klein2024b}. This framework, conceptually similar to the SCALPELS algorithm \citep[e.g.][]{cameron2021,john2022}, has been shown to robustly filter long-term stellar activity variations whilst preserving planet RV signatures in most cases. Finally, we blindly apply DI to the input CCFs, and subtract the RVs derived from the best-fitting line profiles from the input RVs. For each of the three cases, we compute a GLS periodogram from the activity-filtered RVs and estimate the period of the most prominent peak and its uncertainty using $\chi^{2}$ statistics. In each case, the planet is said to be detected if (1)~the injected orbital period lies within 3$\sigma$ of the recovered one, and (2)~the false alarm probability of the recovered period, computed using \citet{baluev2008}'s criterion, is smaller than 0.1\%. It is worth noting that, without activity filtering, none of the injected planets would be retrieved with this criterion.

\begin{figure*}
    \centering
    \includegraphics[width=\linewidth]{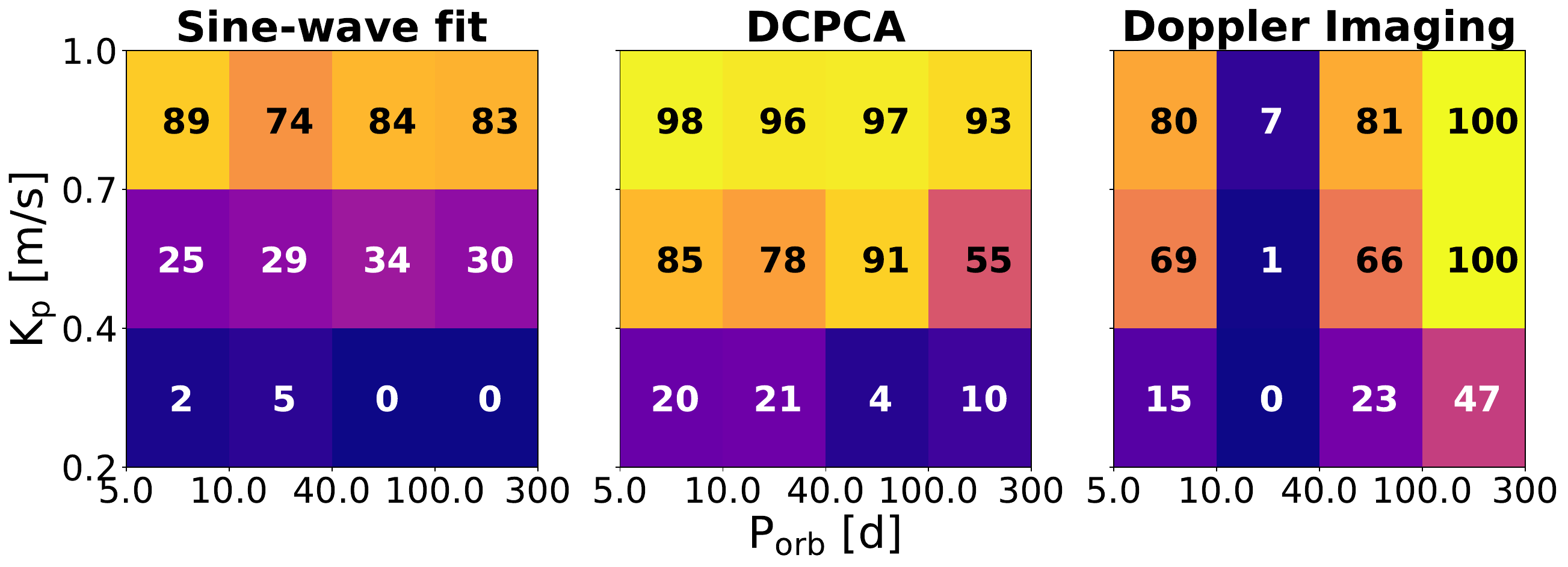}
    \caption{Completeness estimates for the recovery of the 1\,000 planet RV signatures injected in the HARPS-N solar CCFs for different activity filtering techniques (see Sec.~\ref{sssec:inj_rec}). The number and the color of each cell both indicate the completeness of the planet injection-recovery, in percent.}
    \label{fig:inj_rec}
\end{figure*}

The completeness maps of the planet injection-recovery tests for the three activity-filtering techniques are shown in Fig.~\ref{fig:inj_rec}. The sine-wave is too simplistic and only filters part of the stellar activity signal. Therefore, if planets with RV semi-amplitudes larger than $\sim$0.7\,\ms\ are relatively well recovered, the completeness decreases quickly with decreasing RV semi-amplitude. The DCPCA framework is significantly more sensitive to smaller-amplitude planet signatures, with about 87\% of completeness for planet semi-amplitudes larger than 0.4\,\ms. The completeness completely drops for smaller semi-amplitudes, which we attribute to the fact that DCPCA only provides a partial filtering of quasi-periodic stellar activity signals \citep[see Figure 7 of][]{klein2024b}. DI performs exceptionally well with long-period planets (i.e. P$_{\mathrm{orb}} >$100\,d), with the totality of injected planets with semi-amplitude larger than 0.4\,\ms\ and a completeness of 67\% for semi-amplitudes smaller than 0.4\,\ms. On the other hand, almost all planets with orbital periods near the Sun's rotation period and first harmonics are lost in the process, which is not surprising as DI is driven by the modulation of CCF distortions at the star's rotation period. In the DI case, we also note that most of the discarded signals with $P_{\mathrm{orb}} \lesssim 10$\,d and $K_{\mathrm{p}} \gtrsim 0.4 $\,\ms\ have an orbital period between 8.5 and 9.5\,d, which corresponds to the third harmonic of the solar rotation period (see also Fig.~\ref{fig:rv_per}). Nonetheless, this test demonstrates that DI can, in principle, be used as a planet search tool, and complement well other methods in the planet parameter space to which it is sensitive.

\subsection{Joint activity and planet fit}

The previous section has demonstrated the need to incorporate the planet search directly into the DI framework. This framework is first described in Sec.~\ref{sssec:521}, and then applied to recover the three planetary signals listed in Tab.~\ref{tab:planet_prop}.

% \begin{table*}
%     \caption{Best-fitting estimates.}
%     \label{tab:my_label}
%     \centering
%     \begin{tabular}{cccccccccc}
%         \hline
%        Model & Case &  K$_{\mathrm{inj}}$ & P$_{\mathrm{inj}}$ & K$_{\mathrm{est}}$ & P$_{\mathrm{est}}$ & $\sigma_{\mathrm{j}}$ & $\lambda_{\mathrm{e}}$ & $\lambda_{\mathrm{p}}$ & P$_{\mathrm{rot}}$  \\
%        --  & -- &  [\ms] & [d] & [\ms] & [d] & [\ms] & [d] & -- & [d] \\
%          \hline
%         & Planet 1 & 1.0 & 6.0 & 0.59$^{+0.04}_{-0.05}$ & 6.000\,$\pm$\,0.002 & 0.59\,$\pm$\,0.02 & -- & -- & -- \\
%       D$_{1}$  & Planet 2 & 0.4 & 6.0 & 0.25$^{+0.05}_{-0.06}$ & 6.000\,$\pm$\,0.005 & 0.57\,$\pm$\,0.02 & -- & -- & -- \\
%        & Planet 3 & 0.5 & 100.0 & 0.31$^{+0.04}_{-0.06}$ & 99.8$^{+0.9}_{-1.1}$ & 0.59\,$\pm$\,0.02 & -- & -- & -- \\
%        \hline
       
%          \hline
%     \end{tabular}
% \end{table*}

\subsubsection{Method}\label{sssec:521}

We use the method introduced by \citet{petit2015}, and validated on various cases \citep[e.g.][]{donati2017,yu2017,klein2021,heitzmann2021,klein2022}. We start from a grid of parameters (K$_{\mathrm{p}}$, P$_{\mathrm{orb}}$ and $\phi_{\mathrm{p}}$ in this case). For each set of parameters in the grid, we compute the planet RV signature using Eq.~\ref{eq:planet}, and use this signature to shift the wavelength of each observed CCF. We then apply DI to the shifted CCFs to a given level of entropy, which gives a value $\chi^{2}$ quantifying the goodness of the fit. We repeat this process for all the grid parameters and estimate the best-fit parameters and uncertainties from the resulting $\chi^{2}$ distribution \citep{press1992}. The planet search is performed independently on each of the 23 chunks of Tab.~\ref{tab:list_chunks}. The best estimate of the planet parameters is simply obtained by averaging the best-fit parameters over all chunks and propagating uncertainties.

For comparison, we use one- and two-dimensional Gaussian Processes (GPs) to model stellar activity signals while fitting for the planet RV signature. We use the GP framework of \citet{rajpaul2015} as implemented in the \texttt{pyaneti} software of \citet{barragan2022A}. This framework uses a quasi-periodic GP to represent the activity signal, with covariance kernel between epochs $t_{i}$ and $t_{j}$ given by:

\begin{equation}
    k(t_{i},t_{j}) = A^{2} \exp \left[ - \frac{(t_{j}-t_{i})^{2}}{2 \lambda_{\mathrm{e}}^{2}} - \frac{\sin^{2} \pi (t_{j}-t_{i})/P_{\mathrm{GP}}}{2 \lambda_{\mathrm{p}}^{2}} \right], 
    \label{eq:qp_kernel}
\end{equation}

\noindent
where $A$ is the GP amplitude, and where the GP period $P_{\mathrm{GP}}$, evolution timescale $\lambda_{\mathrm{e}}$ and inverse harmonic complexity $\lambda_{\mathrm{p}}$ are the three hyperparameters of our model. In the one-dimensional case, we model the RVs only, and our model is composed of a planet component computed with Eq.~\ref{eq:planet} (i.e. three free parameters, K$_{\mathrm{p}}$, P$_{\mathrm{orb}}$ and $\phi_{\mathrm{p}}$), a stellar activity signal modelled by a quasi-periodic GP (following Eq.~\ref{eq:qp_kernel}), and a white noise term, defined as the quadratic sum of the photon noise and an additional uncorrelated jitter term $\sigma_{\mathrm{j}}$, to absorb variations not accounted for by the GP. In the two-dimensional case, the RVs are modelled jointly with the CCF FWHM, known to be a reliable proxy of the solar activity RV signals \citep{klein2024b}. As described in \citet{rajpaul2015}, the RV and FWHM activity signals are modelled as linear combinations of a latent variable (typically the square of the photometric flux) and its first temporal derivative, following the FF' framework of \citet{aigrain2012}. This latent variable is modelled as a GP with the quasi-periodic covariance kernel of Eq.~\ref{eq:qp_kernel}.

\subsubsection{Results}\label{sssec:522}

We proceed to recover the three planetary signals listed in Tab.~\ref{tab:planet_prop}. Recovering the planet RV semi-amplitude and orbital period and phase is challenging, even for multi-dimensional GPs. However, as shown in Section~\ref{ssec:5.1}, the blind CCF modelling with DI has enabled us to accurately estimate the planet's orbital period and phase. We thus adopt Gaussian priors for these two parameters, using the best-fit values and uncertainties reported in Tab.~\ref{tab:D1_results}. For Doppler Imaging, the grids on P$_{\mathrm{orb}}$ and $\phi_{\mathrm{p}}$ are centered on the mean prior values and extend up to 10 standard deviations. A Uniform prior between 0 and 5\,\ms\ is adopted for K$_{\mathrm{p}}$. Non-informative priors are adopted for the other parameters of the model.

{\renewcommand{\arraystretch}{1.1}
\begin{table*}
    \centering
    \caption{Best-fit estimates of the RV semi-amplitude of the planets described in Tab.~\ref{tab:planet_prop} and recovered using one- and two-dimensional GPs and DI. All semi-amplitudes are given in \ms\ with 1$\sigma$ uncertainties.}
    \label{tab:results_D2}
    \begin{tabular}{cccccccccccc}
    \hline
         &  \multicolumn{3}{c}{Planet 1 (P$_{\mathrm{orb}}$=6d, K$_{\mathrm{p}}$=1\ms )}  &  & \multicolumn{3}{c}{Planet 2 (P$_{\mathrm{orb}}$=6d, K$_{\mathrm{p}}$=0.4\ms )}  & &  \multicolumn{3}{c}{Planet 3 (P$_{\mathrm{orb}}$=100d, K$_{\mathrm{p}}$=0.5\ms )}     \\
   \cline{2-4} \cline{6-8} \cline{10-12}
  Orbital phase   & 0.0 & 0.33 & 0.7 &  & 0.0 & 0.33 & 0.7 &  & 0.0 & 0.33 & 0.7 \\
       \hline
1D GP & 0.95$_{-0.07}^{+0.05}$ & 0.95$_{-0.07}^{+0.06}$ &       1.01$_{-0.06}^{+0.05}$ &  &  0.37$\pm$0.05 & 0.37$\pm$0.05 &       0.43$\pm$0.05 &  &  0.46$\pm$0.25 & 0.43$_{-0.25}^{+0.26}$ &       0.36$_{-0.23}^{+0.27}$ \\
2D GP & 0.96$_{-0.07}^{+0.08}$ & 0.94$\pm$ 0.06 & 1.05$\pm$0.06 &  &       0.37$\pm$0.06 &  0.35$\pm$0.06 & 0.46$\pm$0.06 &  & 0.66$\pm$0.07 & 0.54$\pm$0.06 & 0.36$\pm$0.06 \\
DI & 1.02$\pm$0.05 & 1.03$\pm$0.06 & 0.95$\pm$0.05 & & 0.39$\pm$0.05 & 0.35$\pm$0.05 & 0.41$\pm$0.05 &  & 0.45$\pm$0.05 &  0.48$\pm$0.05 & 0.47$\pm$0.05 \\
\hline
    \end{tabular}
\end{table*}
}

\begin{figure}
    \centering
    \includegraphics[width=\linewidth]{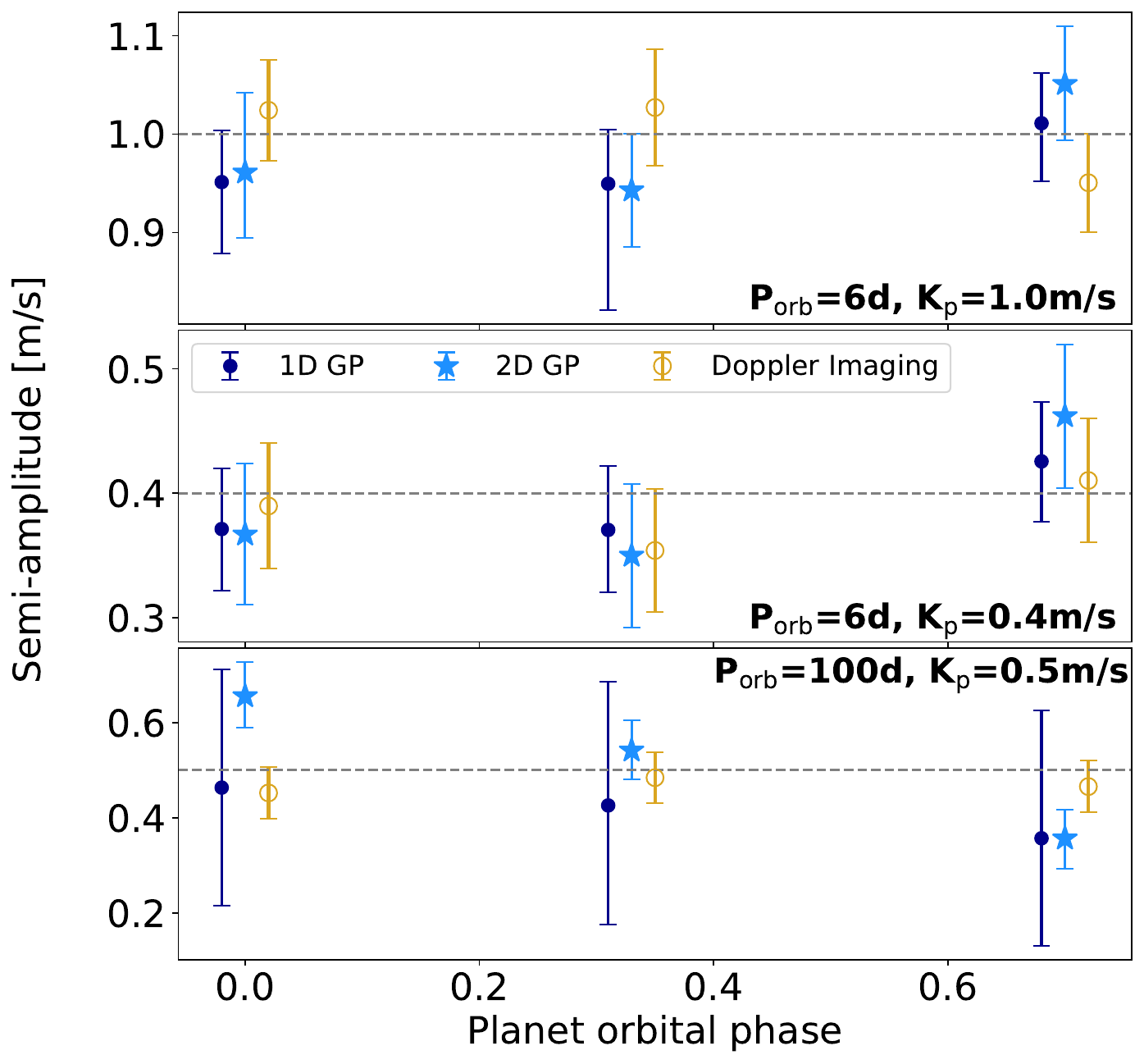}
    \caption{Best estimates of the planet RV semi-amplitude for the three different planets listed in Tab.~\ref{tab:planet_prop} and retrieved using three different stellar activity models. In each panel, the injected RV semi-amplitude is indicated by the horizontal dashed line.}
    \label{fig:mass-Estim}
\end{figure}

The planet RV semi-amplitudes recovered by our different models are given in Tab.~\ref{tab:results_D2} and shown in Fig.~\ref{fig:mass-Estim}. We find that the recovered planet RV semi-amplitudes lie systematically within $\sim$2$\sigma$ of the injected value, for each injected planet and activity modelling technique. Additionally, we do not identify any systemtic trend between the recovered semi-amplitude and the injected planet orbital phase. As already shown in \citet{klein2024b}, the one-dimensional GP is only able to provide a marginal $\sim$1.8$\sigma$ detection of the 100-d-period planet, due to long-term activity evolution. The two-dimensional GP provides significantly more precise semi-amplitude estimates for Planet~3. However, as we change the planet orbital phase, we do observe significant variations of $\pm$2 $\sigma$ of the recovered value with respect to the injected one. In contrast, the semi-amplitudes estimates from DI are systemtically consistent with the injected values at 1$\sigma$, with errorbars comparable to those obtained with the two-dimensional GP.

\section{Discussion and conclusions}\label{sec:conclusion}

In this proof-of-concept study, we assessed the ability of traditional Doppler Imaging to model the magnetic activity and search for low-mass planets around of Sun-like stars. We divided two years of Sun-as-a-star observations, collected with the high-precision spectrograph HARPS-N between 2022 and 2024, into 27-day-long chunks of data, and modelled the time series of CCFs within each chunk with DI. The output relative brightness topologies, shown in orthographic projection in Fig.~\ref{fig:best_maps} and in rectangular view in Fig.~\ref{fig:rect_maps}, exhibit typical filling factor of 0.3\%, and angular resolution of about 36$^{\circ}$, and vary significantly from one rotation cycle to the next, as expected from the fast-evolving solar activity. We found that the RVs derived from the best-fitting line profiles match well the HARPS-N solar RVs, with residuals of 0.58\,\ms\ RMS (see Fig.~\ref{fig:rv_fit}), similar to the level obtained with Gaussian Process modelling \citep{klein2024b}, and no sign of modulation at the solar rotation period and harmonics.

\subsection{Accuracy of the brightness maps}

We found that the brightness maps obtained with DI match well the large-scale structures of contemporaneous SDO Dopplergrams (when modelled with spherical harmonics of degree $\ell_{\mathrm{max}}$\,=\,5), provided that the absolute value of the RV at the rotational phase of comparison is larger than $\sim$2\,\ms. Smaller RVs reflect CCF distortions that are barely larger than the noise and are generally less well recovered with Doppler Imaging. The DI brightness maps match significantly less well the SDO/HMI intensitygrams (probably because the main cause of CCF distortions is the inhibition of convective blueshift rather than brightness contrast effects), and the magnetograms, which exhibit structures with higher degree of complexity.

The match between Doppler images and SDO Dopplergrams could bring constraints on the location of stellar active regions for the characterisation of exoplanet atmospheres with space-based transmission spectroscopy. Stellar activity is currently one of the main limitations in the characterisation of the atmosphere of rocky planets with the James Webb Space Telescope \citep[e.g.][]{rackham2018,rackham2019,tondreau2024,radica2024}. Knowing the position of the largest active regions beforehand would help refine stellar activity models in exoplanet atmosphere retrievals. For the time being, we have shown that Doppler Imaging is able to pick up spatially-resolved information that correlates moderately well with SDO/HMI Dopplergrams. However, more work is needed to understand what physical quantities are extracted with DI and how to convert DI maps into physical constraints for planet atmosphere retrievals. Realistic simulated spectra \citep[e.g.][]{sowmya2023,smitha2025} or disc-resolved high-resolution solar spectra, like those that the POET instrument will collect \citep{santos2025}, will help improve the accuracy of the DI modelling.

% pick up information that correlates well with the HMI Dopplergrams

% description of the large-scale variations at the stellar surface  }

% By providing a relatively accurate description of the large-scale brightness variations at the stellar surface, Doppler imaging maps could be used to refine stellar activity models in exoplanet atmosphere retrievels. This will work best for active stars (e.g. pre-main-sequence stars, active M dwarfs) and will require contemporaneous high-resolution spectra, preferrably at visible wavelengths, were the effects of activity are the most identifiable. However, given the degeneracy between the temperature and filling factor of the active regions reconstructed with DI, we caution that converting DI maps into physical constraints for planet atmosphere retrievals requires dedicated studies. 

\subsection{Towards an operational version of Doppler Imaging for Sun-like stars}

As shown in Section~\ref{sec:planets}, DI is a reliable algorithm for searching for long-period low-mass planet signatures, with accurate retrieval of the planet's orbital period and phase. However, blindly applying DI to a dataset without jointly fitting for planet signature will result in largely under-estimated planet masses. In contrast, when the DI activity modelling is jointly performed with a planet fit, the planet mass estimates are as accurate as those obtained with multidimensional Gaussian process regression, which is arguably the current state of the art activity filtering algorithm. These promising results suggest that Doppler Imaging could become a reliable alternative to traditional RV-based activity modelling techniques in the search for long-period low-mass planets, provided that a number of limitations, detailed below, are addressed.

Performing the Doppler Imaging inference and planet search in a Bayesian framework is a natural next step motivated by the results of this study. However, this requires the DI inverse process to be made significantly faster and more scalable than most of the DI codes used in the literature, including the one used in this study. Recently, \citet{luger2021} made significant progress in this direction by developing a linear, scalable and open-source version of a Doppler Imaging code\footnote{Available through the \texttt{starry} python package, \url{https://starry.readthedocs.io/en/latest/notebooks/DopplerImaging_Intro/}}. Though the linearisation of the DI process may make it difficult to incorporate some physical processes like differential rotation or convective blueshift inhibition, this code is currently our best hope for a Bayesian mapping of stellar surfaces from high-resolution spectra. In fact, Information Field Theory, which describes a physical field using Bayesian statistics \citep[see][]{ensslin2009,ensslin2011}, provides a flexible framework fully adapted to Doppler Imaging, bypassing the constraints of regularisation (e.g. maximum entropy, Tikhonov).

Another fundamental limitation of current DI algorithms is that they generally do not account for the evolution of stellar activity with time. This is a problem for the search of long-period planets, which requires DI to be performed on long baselines, and for stars with fast-evolving activity, like the Sun. Describing the temporal evolution of the coefficients of the stellar surface map with Gaussian processes is one of the most promising way of addressing this issues. This, in fact, has been recently incorporated in light curve modelling \citep{luger2021b} and for Zeeman-Doppler Imaging \citep{finociety2022}. However, this implementation carries the risk of adding further degeneracies or assumptions to the inversion process (e.g. accounting for the different evolution time scales of active regions of different size, latitutes and depth). For Sun-like stars in particular, the rapidly evolving activity might be difficult to capture with Gaussian Processes.

Finally, incorporating more physically accurate descriptions of the stellar surfaces is one of the most exciting long-term perspective of Doppler imaging, taking advantage of the unrivalled precision of new-generation spectrographs. Inverting the Sun's CCF distortions into a velocity field rather than into a relative brightness distribution will likely improve the fit quality, especially in the line cores where the largest deviations are observed (see Fig.~\ref{fig:best_profiles_1} and~\ref{fig:best_profiles_2}). This should also arguably improve the match with SDO Dopplergrams. To go further, modelling the full spectra of Sun-like stars, or, in the first instance, CCFs computed using tailored line masks, will allow us to separate different spectral contributions of stellar activity. For example, whereas contrast-induced distortions of stellar line profiles, dominated by spots, have a well-known chromatic dependence, the inhibition of the convective blueshift in faculae depends on the line formation temperature \citep{almoulla2022a,rescigno2025}. These spectral differences could, in principle, enable one to produce an intensitygram and a Dopplergram from the same time series of spectra, provided that the SNR of the spectra is high enough. Similarly, new-generation near-infrared (nIR) spectra of M dwarfs and pre-main-sequence stars are primarily sensitive to two effects of stellar activity: the spot-induced contrast and the Zeeman splitting in magnetically sensitive lines \citep{reiners2013,klein2021,donati2023}. Accounting for the different magnetic sensitivities (i.e. Land\'e factors) of the spectral lines used in the DI inversion could, in principle, enable one to separate the Zeeman-induced distortions from contrast effects, provided that enough spectral lines with accurate Land\'e factors are available in the nIR. Lastly, other causes of stellar variability like supergranulation \citep[][]{rincon2018,lakeland2024,osullivan2024,osullivan2025}, remain poorly understood and might, in fact, represent an intrinsic limitation to the accuracy of Doppler Imaging maps.

\section*{Acknowledgements}

We thank an anonymous referee for valuable comments that improved the manuscript. This publication is part of a project that has received funding from the European Research Council (ERC) under the European Union’s Horizon 2020 research and innovation program (Grant agreement No. 865624). KA acknowledges support from the Swiss National Science Foundation (SNSF) under the Postdoc Mobility grant P500PT\_230225. ACC acknowledges support from STFC consolidated grant number ST/V000861/1 and UKRI/ERC Synergy Grant EP/Z000181/1 (REVEAL). A.M. acknowledges funding from a UKRI Future Leader Fellowship, grant number MR/X033244/1 and a UK Science and Technology Facilities Council (STFC) small grant ST/Y002334/1.

This study is based on observations made with the Italian Telescopio Nazionale Galileo (TNG) operated on the island of La Palma by the Fundación Galileo Galilei of the INAF (Istituto Nazionale di Astrofisica) at the Spanish Observatorio del Roque de los Muchachos of the Instituto de Astrofisica de Canarias. We thank the HARPS-N solar team and TNG staff for processing the solar data and maintaining the solar telescope.

% The HARPS-N project has been funded by the Prodex Program of the Swiss Space Office (SSO), the Harvard University Origins of Life Initiative (HUOLI), the Scottish Universities Physics Alliance (SUPA), the University of Geneva, the Smithsonian Astrophysical Observatory (SAO), and the Italian National Astrophysical Institute (INAF), the University of St Andrews, Queen’s University Belfast, and the University of Edinburgh.
% Based on observations made with the
% Italian Telescopio Nazionale Galileo (TNG) operated on the island
% of La Palma by the Fundación Galileo Galilei of the INAF (Istituto
% Nazionale di Astrofisica) at the Spanish Observatorio del Roque de
% los Muchachos of the Instituto de Astrofisica de Canarias

%%%%%%%%%%%%%%%%%%%%%%%%%%%%%%%%%%%%%%%%%%%%%%%% %%
\section*{Data Availability}

The HARPS-N solar data used in this study can be accessed on the Data Analysis Center
for Exoplanet (DACE), either through the web-platform: \url{//dace.unige.ch/sun/} or through the DACE \texttt{python} API. The data will be extensively described in Dumusque et al., in prep. The SDO/HMI images are publicly available at \url{https://sdo.gsfc.nasa.gov/data/}.



\bibliographystyle{mnras}
\bibliography{biblio} % if your bibtex file is called example.bib

%\begin{thebibliography}{99}
%\end{thebibliography}

%%%%%%%%%%%%%%%%%%%%%%%%%%%%%%%%%%%%%%%%%%%%%%%%%%

%%%%%%%%%%%%%%%%% APPENDICES %%%%%%%%%%%%%%%%%%%%%

\appendix

\section{Doppler Imaging modelling}

% \begin{figure}
%     \centering
%     \includegraphics[width=\linewidth]{Figures/profiles.pdf}
%     \caption{Time series of the first 10 observed (thin black lines) and best-fit (thick red lines) mean-subtracted line profiles in chunks 14 and 15 (see Tab.~\ref{tab:list_chunks}). The $\pm$1$\sigma$ uncertainties on the line continuum is indicated at the right-hand side of each profile. }
%     \label{fig:best_profiles}
% \end{figure}

\begin{figure*}
    \centering
    \includegraphics[width=0.95\linewidth]{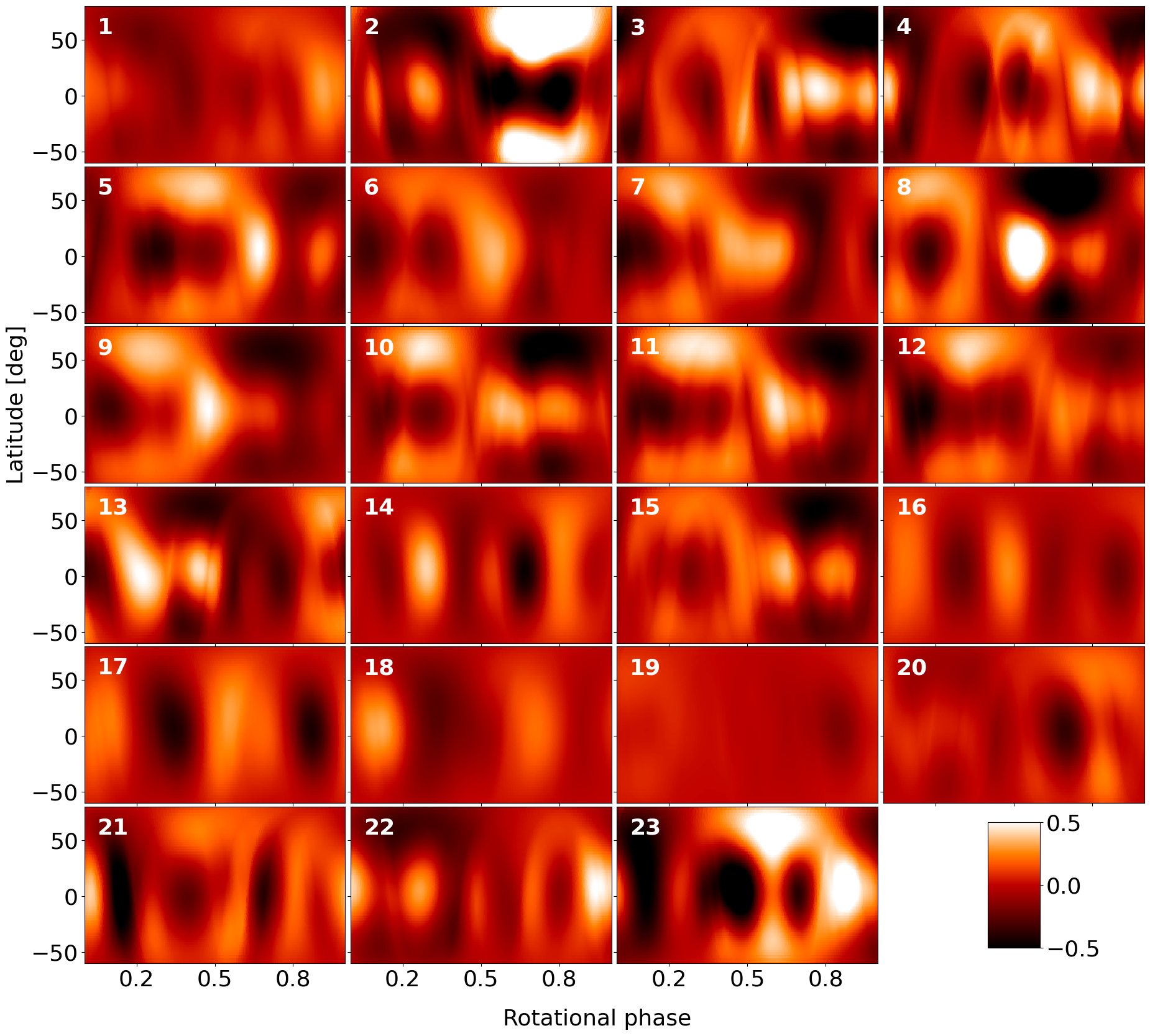}
    \caption{Best-fit relative brightness maps for the 23 chunks listed in Tab.~\ref{tab:list_chunks}. As in Fig.~\ref{fig:best_maps}, the color scale depicts the logarithm of the relative surface brightness, in percent. Low latitudes are truncated given our assumed inclination of 80$^{\circ}$.}
    \label{fig:rect_maps}
\end{figure*}

\begin{figure*}
    \centering
    \includegraphics[width=\linewidth]{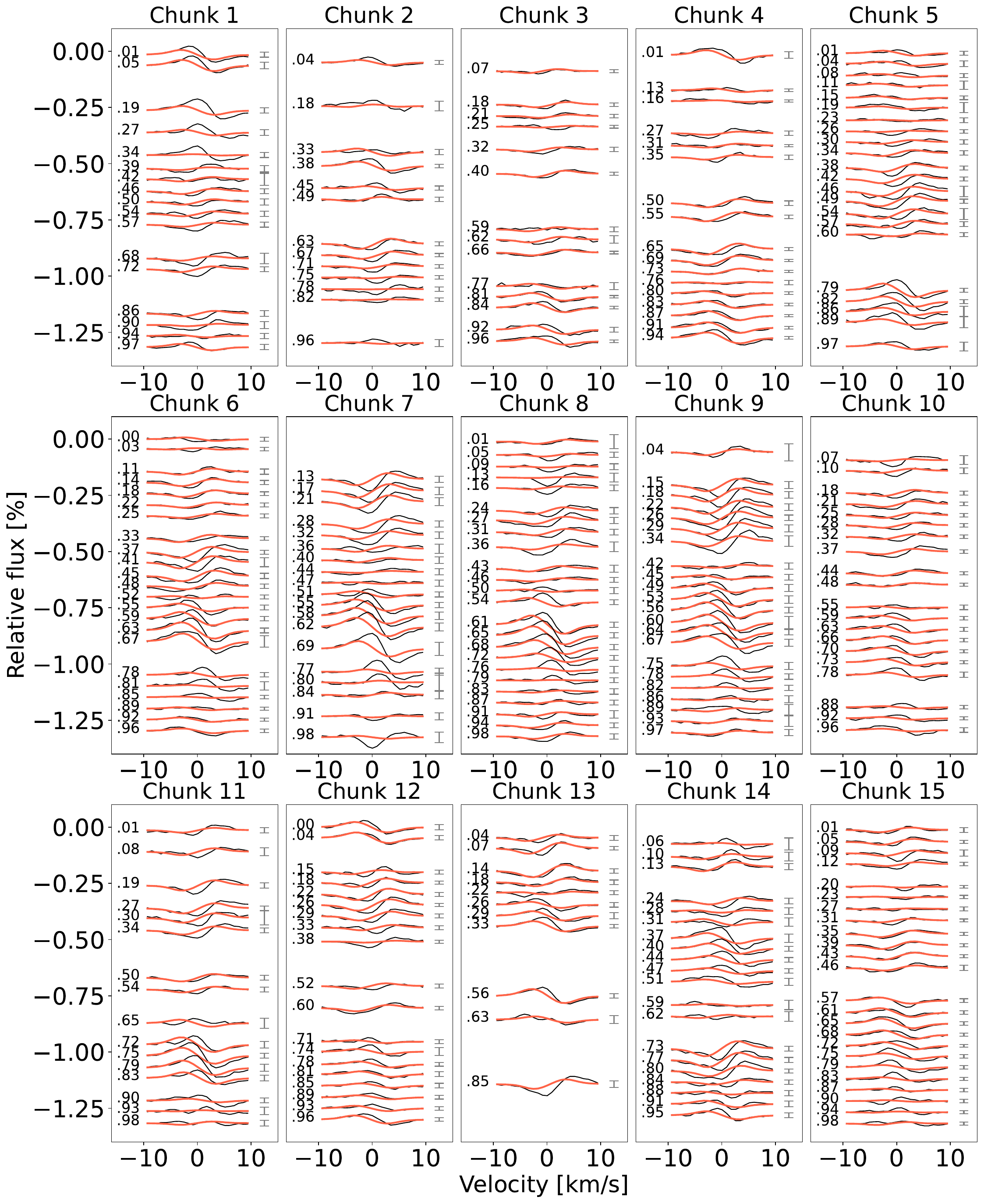}
\caption{Time series of all observed (thin black lines) and best-fit (thick red lines) mean-subtracted line profiles in the first 15 chunks of Tab.~\ref{tab:list_chunks}. The $\pm$1$\sigma$ uncertainties on the line continuum is indicated on the right-hand side of each profile. For each chunk, the line profiles are stacked vertically according to the Sun's rotational phase, computed with Eq.~\ref{eq:phase}, and indicated on the left-hand side of each observation. Note that, in the rare event that two line profiles from the same data chunk are obtained at the same rotational phase (e.g., following a complete solar rotation), only one profile is shown in the figure for clarity (which happened in Chunk 8 at phase 0.36, and in Chunk 15 at phase 0.35).}   
    \label{fig:best_profiles_1}
    \end{figure*}

\begin{figure*}
    \centering
    \includegraphics[width=\linewidth]{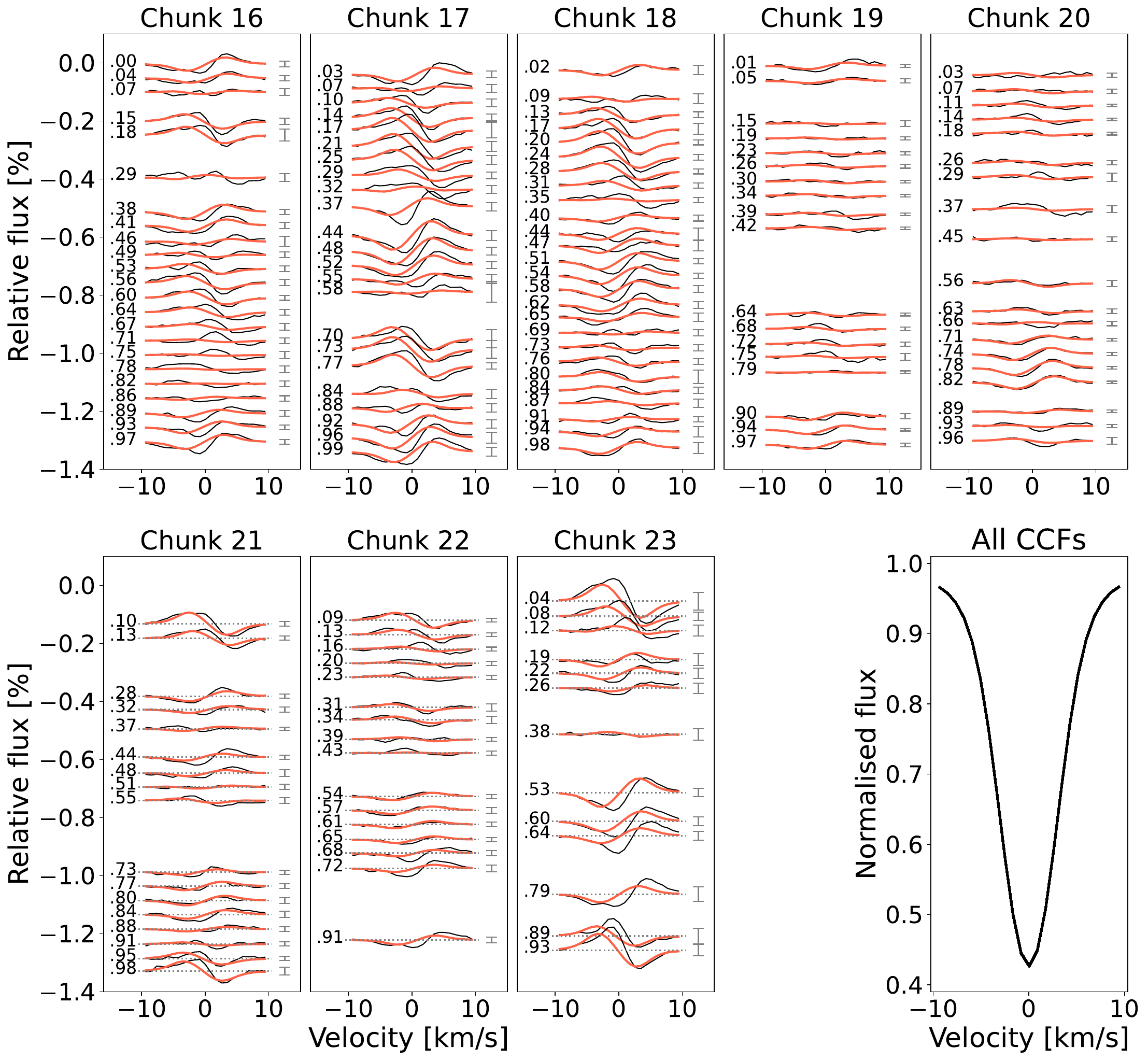}
\caption{Same as Fig.~\ref{fig:best_profiles_1} for chunks 16 to 23. As a reference, we also superimposed all continuum-normalised CCFs in the bottom right panel. To make the figure clearer, one observation (Chunk 18 at phase 0.35) is not shown in this figure.}
    \label{fig:best_profiles_2}
    \end{figure*}

\begin{figure}
    \centering
    \includegraphics[width=\linewidth]{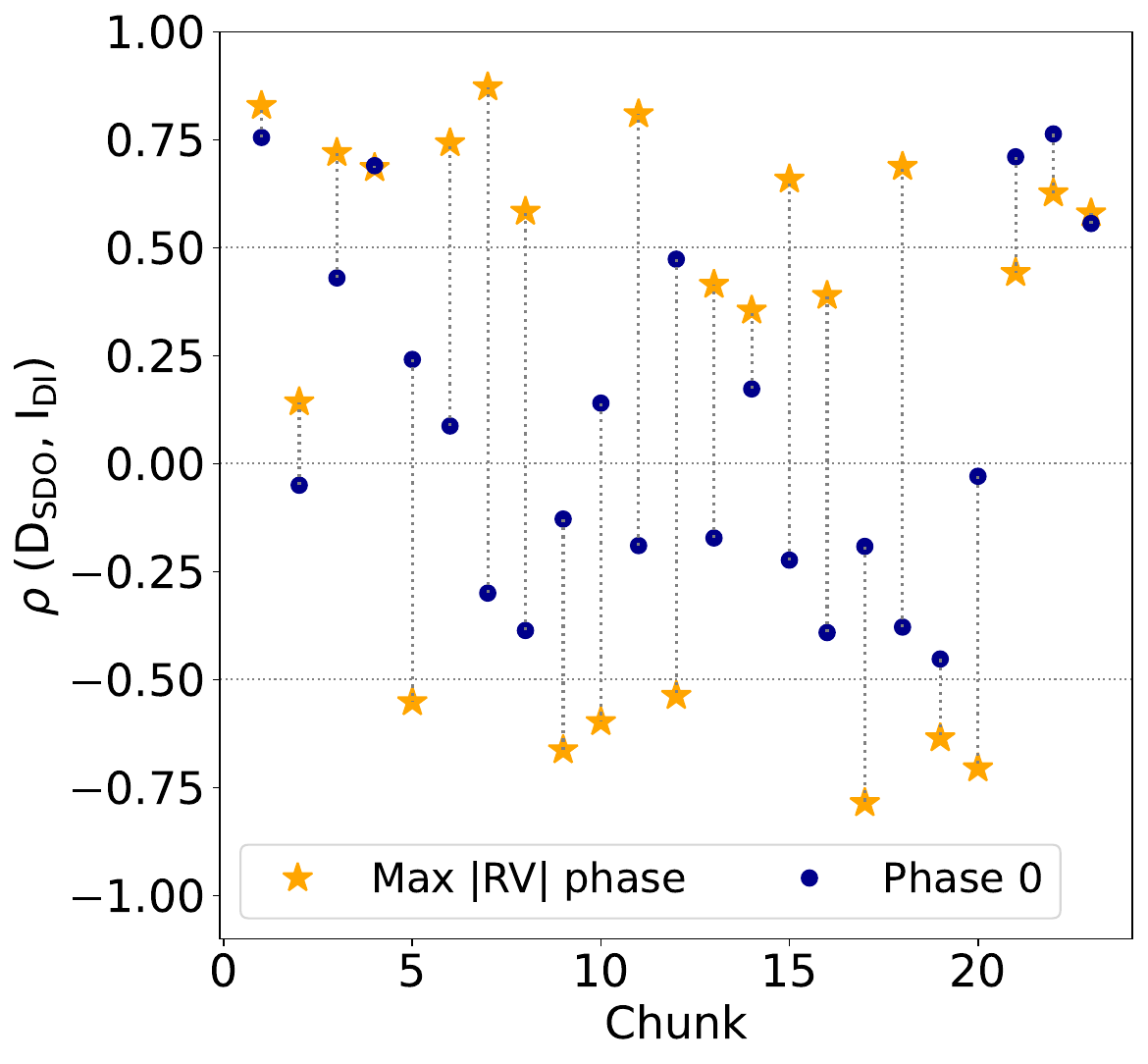}
    \caption{Pearson correlation coefficients between SDO large-scale Dopplergrams (D$_{\mathrm{SDO}}$) and DI brightness maps (I$_{\mathrm{DI}}$) for each chunk at rotational phase 0 (blue dots) and the the phase with maximum absolute RV value (orange stars).}
    \label{fig:all_corr_coef}
\end{figure}

\begin{figure}
    \centering
    \includegraphics[width=\linewidth]{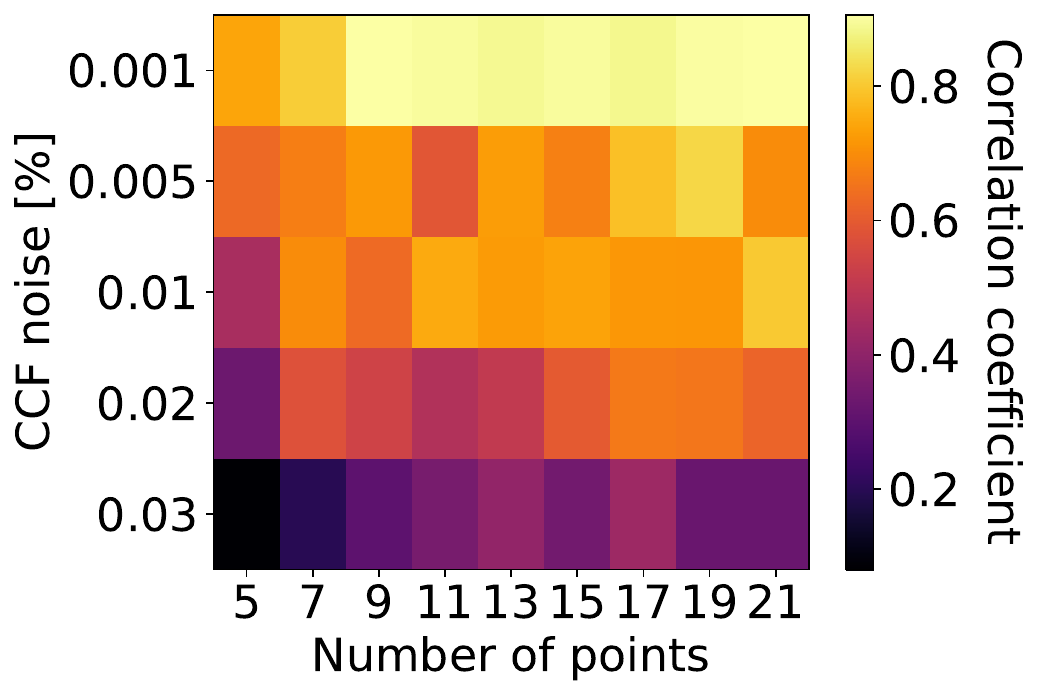}
    \caption{Pearson correlation coefficients between an input brightness distribution and the Doppler maps recovered for different noise levels and number of observations, as described in Sec.~\ref{ssec:4.3}.}
    \label{fig:correl_map_noise}
\end{figure}

In this appendix, we give additional details on the Doppler Imaging reconstruction of the surface distribution of the Sun's brightness from HARPS-N spectra. In Fig.~\ref{fig:rect_maps}, we show a flattened polar view of the best-fitting brightness distributions for all the chunks. We also show the best-fit line profiles in Fig.~\ref{fig:best_profiles_1} and \ref{fig:best_profiles_2}. In fig.~\ref{fig:all_corr_coef}, we show the correlation coefficients between SDO large-scale Dopplergrams and the DI relative brightness maps for each chunk at rotation phase 0 and at the phase with maximum absolute RV value. Finally, as a complement to Sec.~\ref{ssec:4.3}, we show in Fig.~\ref{fig:correl_map_noise} the evolution Pearson correlation coefficient between an input brightness distribution and the DI map recovered, as a function of the number of points and noise level.

% \begin{figure*}
%     \centering
%     \includegraphics[width=0.95\linewidth]{Figures/polar_all.png}
%     \caption{Best-fitting brightness distributions of the Sun for each chunk listed in Tab~\ref{tab:list_chunks}. Each map is a flattened polar view of the star: the pole is located at the center of the plot and concentric circles indicate the Sun's equator (solid line) and -30$^{\circ}$, 30$^{\circ}$ and 60$^{\circ}$ parallels. The color scale depict the logarithm of the relative surface brightness. The radial ticks around the star mark the rotational phases of the observations, as defined in Eq.~\ref{eq:phase}.}
%     \label{fig:polar_maps}
% \end{figure*}

% \section{Goodness of fit}

% \subsection{Best fitting line profiles}

% \begin{figure}
%     \centering
%     \includegraphics[width=\linewidth]{Figures/prediction_10.pdf}    \\ 
%     \includegraphics[width=\linewidth]{Figures/prediction_15.pdf}  \\ 
%     \includegraphics[width=\linewidth]{Figures/prediction_24.pdf} 
%     \caption{Best fitting Doppler imaging profiles for chunk 10 (top panel), 15 (middle panel) and 24 (bottom panel; see the definition of the chunks in Tab.~\ref{tab:list_chunks} and Fig.~\ref{fig:chunks}). In each panel, the observed CCFs best-fitting synthetic profiles are shown in thin black lines and thick orange lines, respectively. The solar rotation phases, computed in Sec.~\ref{ssec:chunks}, are given at the left-hand side of each profile. On the right-hand side of each profile, we indicate the $\pm$1\,$\sigma$ uncertainty on the profile continuum. For clarity, each profile is mean-subtracted and shifted horizontally and/or vertically.}
%     \label{fig:predictions}    
% \end{figure}

% \subsection{Cross validation}

% \begin{figure}
%     \centering
%     \includegraphics[width=\linewidth]{Figures/chi2_map.pdf}
%     \caption{Caption}
%     \label{fig:chi2_map}
% \end{figure}

%%%%%%%%%%%%%%%%%%%%%%%%%%%%%%%%%%%%%%%%%%%%%%%%%%

% Don't change these lines
\bsp	% typesetting comment
\label{lastpage}
\end{document}